\documentclass[journal]{IEEEtran}
\usepackage[latin9]{inputenc}
\usepackage{amsmath}
\usepackage{amssymb}
\usepackage{graphicx}
\usepackage{esint}
\usepackage{algorithm}
\usepackage{algpseudocode}
\usepackage{pbalance}
\algtext*{EndFor}
\algtext*{EndIf}
\algtext*{EndProcedure}

\usepackage{gensymb}
\usepackage{array} 
\usepackage{xurl} 
\usepackage[colorlinks=true, allcolors=blue]{hyperref}
\makeatletter
\DeclareTextSymbolDefault{\textquotedbl}{T1}
\providecommand{\tabularnewline}{\\}

\usepackage{array}
\usepackage{url}
\usepackage{multirow}
\usepackage{cuted}

\DeclareTextSymbolDefault{\textquotedbl}{T1}

\IEEEoverridecommandlockouts
\usepackage{cite}
\usepackage{amsfonts}
\usepackage{textcomp}
\usepackage{xcolor}
\def\BibTeX{{\rm B\kern-.05em{\sc i\kern-.025em b}\kern-.08em
		T\kern-.1667em\lower.7ex\hbox{E}\kern-.125emX}}

\makeatother

\begin{document}
\title{A Unified Approach to Human-Scale Blockage and Scattering Analysis in Sub-THz Propagation With Application to RF Sensing

\thanks{Manuscript received xx xxx xxxx. (Stefano Savazzi and Fabio Paonessa are co-first authors.) (Corresponding author: Stefano Savazzi.)\protect \\
Authors are with National Research Council of Italy, Institute of Electronics, Information and Telecommunications Engineering.\protect \\
Funded by the European Union. Views and opinions expressed are however those of the author(s) only and do not necessarily reflect those of the European Union or European Innovation Council and SMEs Executive Agency (EISMEA). Neither the European Union nor the granting authority can be held responsible for them. Grant Agreement No: 101099491 (project HOLDEN).\protect \\
The measurement setup has been partially developed in the framework of SoBigData.it (Smart Radio Environment). SoBigData.it receives funding from European Union ---NextGenerationEU-National Recovery and Resilience Plan (Piano Nazionale di Ripresa e Resilienza, PNRR) ---Project: ``SoBigData.it ---Strengthening the Italian RI for Social Mining and Big Data Analytics'' ---Prot. IR0000013 ---Avviso n. 3264 del 28/12/2021.\protect}}

\author{Stefano~Savazzi,~\IEEEmembership{Senior~Member,~IEEE,}
Fabio~Paonessa,~\IEEEmembership{Member,~IEEE,}
Sanaz~Kianoush,~\IEEEmembership{Member,~IEEE,}
Alessandro~Nordio,~\IEEEmembership{Member,~IEEE,}
and~Giuseppe~Virone,~\IEEEmembership{Senior~Member,~IEEE}}

\maketitle
\begin{abstract}
RF sensing exploits phase-sensitive measurements of stray electromagnetic (EM) fields from wireless devices across various frequency bands to detect EM blockage and to reconstruct and map the surrounding environment in 2D/3D. Although blockage effects caused by objects or human motion are well-studied in ISM bands and frequencies up to 60~GHz, there is a significant lack of research for frequencies above 100~GHz. \textcolor{black}{The paper proposes a unified signal processing framework for RF sensing in the sub-THz D-band (105--175~GHz), explicitly integrating EM blockage and scattering as a single process through the birth-death dynamics of multipath components (MPCs).} 
\textcolor{black}{The framework extracts, associates, and classifies MPCs from angle-delay measurements using statistically grounded detection and classification, enabling human-scale sensing from a single radio link.}
The modeling and classification of MPCs, along with large-scale EM parameters, are demonstrated through an indoor measurement campaign using multiple test targets. \textcolor{black}{Experimental results show that newly formed, attenuated, and suppressed MPCs can be reliably identified with millimeter-scale delay resolution. Static object localization achieves average positioning errors of $8-20$~cm depending on range and material, while passive human localization yields errors of $12-17$~cm at $0.5$~m and $26-30$~cm at $2$~m, respectively.}
\textcolor{black}{The proposed framework demonstrates that accurate sensing and localization are feasible at sub-THz frequencies using a single link.} 
\end{abstract}


\section{Introduction}
\label{sec:intro}

\begin{figure}
    \centering \includegraphics[scale=0.64]{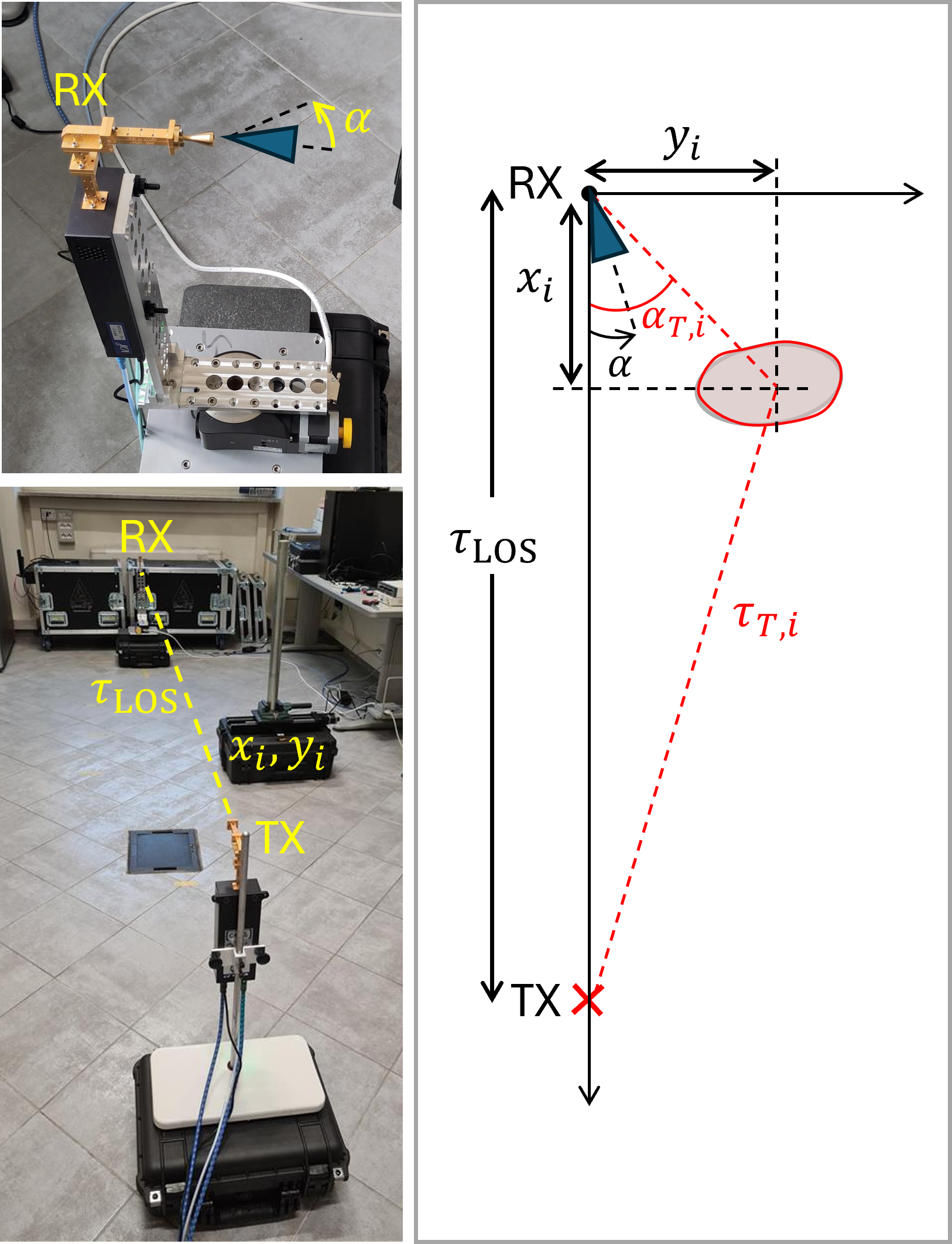} \protect\caption{\label{fig:intro2} \textcolor{black}{Measurement setup and reference system. Top-left: RX mounted on the rotating stage with horn and frequency extender. Bottom-left: fixed TX (near), rotating RX (far), and vertical cylindrical target mounted on the linear stage. Right: top-view sketch. The coordinate origin is at RX. The $x$-axis connects TX and RX, and the $y$-axis lies on the floor plane. For event $i$, $x_{i}$ and $y_{i}$ indicate the target coordinates projected on the floor.} $\alpha$ is the RX antenna orientation angle, w.r.t the $x$ axis. $\alpha_{T,i}$ and $\tau_{T,i}$ are the Angle of Arrival (AoA) and Travel Distance (TD) of the target w.r.t. the RX.}
\end{figure}

Radio Frequency (RF) sensing consists of a set of opportunistic techniques capable of detecting, locating, and tracking people in a monitored area covered by ambient RF signals \cite{wilson-2010}. Using the \emph{Integrated Sensing and Communication} (ISAC) paradigm~\cite{isac,savazzi-1}, the technology can transform each RF node of a wireless network that covers the monitored area into a \emph{virtual sensor}, thus implementing both communication and sensing operations at the same time. RF sensing methods typically employ frequencies in the unlicensed 2.4--5~GHz ISM bands~\cite{shit-2019}, and up to 30--60~GHz~\cite{ojeda-2022, radar60} with  wavelengths ranging from fractions of a centimeter to few centimeters. Such wavelengths are highly subject to multipath effects, which severely limit the sensing accuracy, thus reducing their wide acceptance.

Millimeter-scale wavelengths (mmWave) corresponding to frequencies in the 30--300~GHz range are particularly well suited for high-resolution body occupancy detection and vision applications, as they experience significantly reduced multipath effects, although heavily affected by obstacles blockage. For this reason, the diffusion of THz transceivers for 6G and beyond communication systems~\cite{thz} is expected to pave the way towards novel joint communication, sensing, and computing paradigms. \textcolor{black}{However, the shift towards sub-THz bands requires new signal processing and modeling tools capable of capturing the fundamentally different propagation mechanisms emerging at these wavelengths.}

Most of the sensing and vision techniques proposed for localization and behavior recognition~\cite{CSI}, such as Bayesian filtering, tomography, and holography~\cite{kat,holog}, require a detailed understanding of the effects of human/object blockage and its physical/electromagnetic (EM) properties \cite{fall}. Although body effects are well understood considering the classical ISM bands~\cite{ramp,scatt,mohamed-2017} and up to 30--60~GHz~\cite{26G,bloc}, body blockage and micromobility effects at frequencies beyond 100~GHz have not been adequately analyzed to date \cite{access_new}. For example, a multi-band channel initial characterization based on angular delay profiles has been discussed~\cite{tap} while other studies proposed physical and statistical models to characterize the multipath environment \cite{vehicular} or to assess the impact of human blockage on communication performance \cite{propagation_study,compcomm}. However, a small body of knowledge is available with respect to EM and physical characterization of the body-induced fading effects, which is critical in human-scale sensing applications. \textcolor{black}{In particular, existing approaches largely focus on blockage modeling and do not explicitly address the dynamic birth and death behavior of multipath components induced by human-scale occupancy events.}

In this paper, we provide a study targeting the characterization of human-scale blockage and scattering effects in the D-band 105--175~GHz range, often referred to as sub-THz\footnote{Although this interval extends slightly beyond the conventional D-band definition, we will hereafter denote it as the D-band for the sake of simplicity.}. \textcolor{black}{The primary contribution of this work is the development of a unified signal processing and modeling framework for RF sensing in the sub-THz band. In particular, the paper targets the accurate modeling of the complex interactions between sub-THz radio waves and the blockers (objects or human bodies).} Current approaches often treat radio wave attenuation due to absorption and scattering as separate phenomena primarily to simplify the modeling of the obstruction. However, when the wavelength is smaller than the human body size and features, both effects have a significant impact and can be represented as a unified process. 

The presence of a moving object or body in the proximity of a RF link alters the propagation conditions, giving rise to new multipath components (MPCs) while attenuating or suppressing existing ones. Since a single object can act as a blocker and/or a scatterer along the path, a unified approach is required for their analysis. \textcolor{black}{To this end, we introduce a general processing pipeline that explicitly extracts, associates, and classifies MPCs based on their angle-delay features, enabling the systematic modeling of their birth-death dynamics.} Blockage and scattering events are modeled as variations in both the line-of-sight (LOS) and non-line-of-sight (NLOS) channel components. The MPCs associated with object/body occupancy events are classified into four categories: newly generated, attenuated, suppressed, or unchanged. For each classified component, the joint processing of time-of-arrival (ToA), or travel distance (TD), and angle-of-arrival (AoA) allows the characterization of the occupancy event. 

To validate our approach, we designed an automated spatial probing system that leverages computer-controlled positioning and orientation of antennas to precisely measure both the AoA and ToA/TD of the detected MPCs. An indoor measurement campaign was conducted to collect the RF footprints induced by the target at different blocker positions. \textcolor{black}{It is worth noting that the proposed framework is purposely designed to operate with a single radio link, in contrast to classical RF sensing approaches \cite{savazzi-1}, that require multiple links.}

The main contributions of this work are as follows:
\begin{itemize} 
    \item a unified framework for extracting and classifying the MPCs for RF sensing applications and under various object/body occupancy events affecting a single radio link; 
    \item characterization of the birth--death process of MPCs across different occupancy scenarios, including temporal--spatial variations due to unintentional body movements; 
    \item analysis of localization accuracy and precision over different operating bandwidths, with implications for ISAC frameworks;
    \textcolor{black}{\item an extensive indoor measurement campaign in the D-band, investigating blockage and scattering effects from objects and human bodies, complemented by a benchmark comparison with a K-band system ($18.5-26$~GHz).}
\end{itemize}

The remainder of the paper is organized as follows. Section~\ref{sec:setups} presents the measurement setup and the spatial probing system for MPC estimation. Section~\ref{sec:Channel-Impulse-Response} introduces the channel impulse response (CIR) model and the proposed classification of the MPCs. 
Section~\ref{sec:Experimental-results-in} reports the measurement results with representative static objects. Section~\ref{sec:bodysensing} addresses passive body localization and discusses relevant case studies as well as a benchmark analysis between K-band and D-band.


\begin{table*}[tp]
\caption{\label{tab:parameters} \textcolor{black}{Vector network analyzer settings and antenna specifications at D-band}}
\centering
\color{black}
\begin{tabular}{ll ll ll}
\hline\hline
\multicolumn{2}{c}{\textbf{\textcolor{black}{Vector network analyzer settings}}} 
& \multicolumn{2}{c}{\textbf{TX antenna}} 
& \multicolumn{2}{c}{\textbf{RX antenna}} \\
\hline
\textbf{Start--stop frequency} & 105--175~GHz & \textbf{Type}        & Open-ended waveguide          & \textbf{Type}        & Conical horn \\
\textbf{Frequency step}        & 8.75~MHz     & \textbf{FHPBW}       & $115^\circ$   & \textbf{FHPBW}       & $13^\circ$   \\
\textbf{Resolution bandwidth}  & 1~kHz        & \textbf{Polarization} & Horizontal   & \textbf{Polarization} & Horizontal     \\
\textbf{TX power}              & 13~dBm       & \textbf{Gain}        & 6.5~dBi       & \textbf{Gain}        & 21~dBi       \\
\hline\hline
\end{tabular}
\color{black}
\end{table*}

\section{Measurement Setup and Environment \label{sec:setups}}

The measurement sessions were conducted in an indoor laboratory environment equipped with radio frequency (RF) instrumentation. The measurement setup included a vector network analyzer (VNA) and a pair of frequency extender modules operating in the frequency range to 105--175~GHz\cite{vdi-vnax}. Each extender featured a WR-6.5 rectangular waveguide output, connected to a conical horn antenna with a nominal gain of 21~dB and a 3~dB beamwidth of 13$^\circ$.

One node operated as transmitter (TX) and the other one as receiver (RX), forming a bistatic measurement setup. Throughout each experiment, the VNA captured the complex transmission coefficients $S_{21}$ measured across the entire operating frequency band of the extenders (swept continuous-wave signal), under various configurations of TX and RX geometry and with different obstacles placed between them.

The monitored area is a rectangular laboratory room measuring $10\times3.4$~m, with a floor-to-ceiling height of $3.5$~m. The TX and RX nodes (Fig.~\ref{fig:intro2} \textcolor{black}{and Fig.~\ref{example-metal}(a) on top}) were separated by a distance of $3.88$~m. The LOS path was horizontally oriented at a height of $68$~cm above the floor, approximately aligned with the long side of the room. The transmitter was mounted on a fixed support near the center of the room, oriented towards the RX during all measurements. The RX was mounted on a precision motorized rotation stage \cite{pi-l611} with an angular accuracy of approximately $0.02^\circ$. Such   rotation angle is hereafter denoted by $\alpha$, where $\alpha=0^\circ$ corresponds to the RX pointing towards the TX (see Fig.~\ref{fig:intro2}). 

A motorized linear translation stage \cite{pi-pm414}, also shown in Fig.~\ref{fig:intro2}), with a repeatability of 6~$\mu$m and a maximum travel range of 300~mm, was used to position the scattering object i.e. either a $5\times72$~cm metal cylinder or a $5.5\times91$~cm paper cylinder (diameter $\times$ height). 

Different scenarios were implemented, during which $S_{21}$ traces were collected for several RX angles $\alpha$, with and without the presence of the target. The target could be a stationary human subject standing at various locations in the scene or the aforementioned scatterers mounted on the linear stage.  The RX angle $\alpha$, the position of the test target ($x_i,y_i$) were recorded together with the acquired $S_{21}$ frequency-response traces to enable data alignment and subsequent post-processing. To this end, a custom user interface and acquisition software was developed to automatically control the motorized stages, trigger the VNA, and manage the download and storage of the acquired data. It is worth mentioning that each experiment lasted a few tens of minutes.

\section{Channel Impulse Response Characterization\label{sec:Channel-Impulse-Response}}

The acquired frequency-domain traces $S_{21}(f;\alpha)$ were collected for each RX rotation angle $\alpha$ using frequency sweeps over $N_{f}=8001$ points equally spaced in the range 105--175~GHz, with a frequency resolution of $\delta_{f}=8.75$~MHz. The frequency step provides a maximum
unambiguous delay (spatial aliasing limit) of $1/\delta_f=114$~ns
($34.3$~m), which effectively suppresses potential aliasing artifacts at the considered frequencies. The total bandwidth $B=70$~GHz enables a temporal resolution of $\delta\tau=1/B=14.3$~ps, and a corresponding spatial resolution
of about $4.30$~mm in free space. The adopted RF parameters, summarized in Table~\ref{tab:parameters}, make the setup suitable for resolving fine features of the channel and the scattering object, while still ensuring an adequate observable delay span\footnote{The duration of a single frequency sweep varies with the chosen bandwidth $B$ up to a maximum of 8 seconds. A narrower resolution bandwidth will improve dynamic range at the expense of measurement speed.}. \textcolor{black}{The resolution bandwidth of 1~kHz is a trade-off between measurement dynamic range \footnote{LOS contribution is 25 dB higher than noise floor of the $S_{21}$ trace} and total sweep time, considering the overall swept bandwidth $B$ (70~GHz) and the chosen frequency resolution $\delta_{f}$. 
It is worth noting that VNA-based frequency sweep inherently assumes quasi-stationarity of the scene over the sweep duration. This assumption is strictly satisfied for the inanimate targets (Section \ref{sec:Experimental-results-in}). For the human-body sensing case, the subject remained in a standing posture; however, small involuntary micro-motions inevitably occur that induce significant phase variations in the D-band. Therefore, the channel may experience slight intra-sweep temporal variations. Considerations towards real-time sensing are discussed in Section~\ref{subsec:posacc}.}

The traces $S_{21}(f;\alpha)$ are zero-padded and converted to time domain via an Inverse Fast Fourier Transform (IFFT). The resulting angle-delay power profile (ADPP) response $r(\tau,\alpha)$ is then obtained by reordering the individual transformed traces for each RX rotation $\alpha$. This representation effectively reveals dominant components, such as the LOS, ToAs/TDs and AoAs of the multipath reflections, providing the basis for the subsequent analysis of MPCs. Power-delay and/or angle profiles are used in various motion sensing and localization-based applications~\cite{xie-2018}, combined with AoA estimation~\cite{wlan}.

In this work we define an \textit{occupancy event}, denoted by $i$, as the presence (position or Region of Interest, ROI) of an object or body near the radio link. The term refers to the mixed behavior of the target, which acts simultaneously as both a blocker and a scatterer. As shown in Fig.~\ref{fig:intro2}, $(x_{i},y_{i})$ are the object floor-projected coordinates defined in a Cartesian reference frame centered at the RX position, where the $x$-axis lies along the LOS direction, and the $y$-axis is parallel to the floor. The observed ADPP $r(\tau,\alpha)$ of an occupancy event in the ROI $i$ can be expressed as 
\begin{equation}
    r(\tau,\alpha)=h_{i}(\tau,\alpha|\boldsymbol{\gamma}_{i},\boldsymbol{\tau}_{i},\boldsymbol{\alpha}_{i})+n,\label{eq:RX}
\end{equation}
where $h_{i}$ is the Channel Impulse Response (CIR) and $n$ is an additive noise term. The CIR is further defined as the superposition of multiple MPCs, each described by amplitudes $\boldsymbol{\gamma}_{i}$, ToA/TDs $\boldsymbol{\tau}_{i}$, and AoA $\boldsymbol{\alpha}_{i}$ features. 
Compared to the target-free scenario ($i=0$), the presence of an object ($i>0$) introduces additional MPCs and may also enhance, attenuate, or suppress existing ones. Blockage and scattering effects on propagation are thus modeled as \emph{alterations} of selected multipath features. Newly formed, dead, attenuated, or unchanged MPCs are closely related to the position of the blocker $(x_{i},y_{i})$ and they can be exploited to infer properties of the obstructing object, including its shape, orientation, and movements.

In the following, we address the reconstruction of the unknown MPCs $(\boldsymbol{\gamma}_{i},\boldsymbol{\tau}_{i},\boldsymbol{\alpha}_{i})$ associated with the event $i$, and classify them with respect to the reference components observed in the target-free case $(\boldsymbol{\gamma}_{0},\boldsymbol{\tau}_{0},\boldsymbol{\alpha}_{0})$.

\begin{table*}[tp]
    \caption{\label{class} Classification of multipath components}
    \centering
    \begin{tabular}{ll ll}
    \hline\hline
    \multicolumn{2}{c}{\textbf{CA-CFAR}} & 
    \multicolumn{2}{c}{\textbf{MPC classification}} \\
    \hline
    \textbf{Training band size} & $[4,3]$ & \textbf{Newly formed MPC} & $\varrho_{D}=2.5$~mm \\
    \textbf{Guard band size} & $[2,3]$ & \textbf{Unchanged or attenuated MPC} & $\varrho_{A}=2$~dB \\
    \textbf{False alarm probability} & $10^{-2}$ & & \\
    \hline\hline
    \end{tabular}
\end{table*}

\subsection{Angle-of-Arrival, Time-of-Arrival and Multipath Component Feature Extraction}

In the target-free case, the CIR is denoted by the function 
$h_{0}(\tau,\alpha \mid \boldsymbol{\gamma}_{0}, \boldsymbol{\tau}_{0}, \boldsymbol{\alpha}_{0})$, 
represented as a discrete dataset of $K_{0}$ components. These components are described by the vectors of amplitudes, ToAs/TDs, and AoAs:
\begin{equation}
    \mathbf{\boldsymbol{\gamma}}_{0}=[\gamma_{k,0}]_{k=1}^{K_{0}}, \quad
    \mathbf{\mathbf{\boldsymbol{\tau}}}_{0}=[\tau_{k,0}]_{k=1}^{K_{0}}, \quad
    \mathbf{\boldsymbol{\alpha}}_{0}=[\alpha_{k,0}]_{k=1}^{K_{0}}.
\end{equation}
Accordingly, the target-free CIR can be expressed as
\begin{equation}
    h_{0}(\tau,\alpha \mid \boldsymbol{\gamma}_{0}, \boldsymbol{\tau}_{0}, \boldsymbol{\alpha}_{0})
    = \sum_{k=1}^{K_0} \gamma_{k,0} \, \delta(\tau-\tau_{k,0}) \, \delta(\alpha-\alpha_{k,0}),
    \label{eq:empty}
\end{equation}
where $\gamma_{k,0}$, $\tau_{k,0}$, and $\alpha_{k,0}$ denote the amplitude, ToA/TD, and AoA of the $k$-th component, respectively, and $\delta(\cdot)$ is the Dirac delta function.

Following an occupancy event $i$, the presence of the blocker induces angle-delay component perturbations that can be analyzed and isolated from noisy observations of the ADPPs ($\ref{eq:RX}$). For ROI $i$, the CIR is modeled as
\begin{equation}
    \begin{aligned}
    h_{i}(\tau, \alpha \mid \boldsymbol{\gamma}_{i}, \boldsymbol{\tau}_{i}, \boldsymbol{\alpha}_{i})
    &= h_{0}(\tau, \alpha \mid \widetilde{\boldsymbol{\gamma}}_{0}, \boldsymbol{\tau}_{0}, \boldsymbol{\alpha}_{0}) \\
    &\quad + \sum_{q=1}^{\triangle K_{i}} \gamma_{q,i} \, \delta(\tau-\tau_{q,i}) \, \delta(\alpha-\alpha_{q,i})\,,
    \end{aligned}
    \label{eq:CIR}
\end{equation}
and is characterized by an augmented set of perturbed MPCs with cardinality $K_{i}=K_{0}+\triangle K_{i}$:
\begin{equation}
    \begin{aligned}
    \boldsymbol{\gamma}_{i} &= \big[\widetilde{\gamma}_{1,0}=\rho_{1,i}\cdot\gamma_{1,0}, \dots, \widetilde{\gamma}_{K_0,0}, \gamma_{1,i}, \dots, \gamma_{\triangle K_i,i}\big],\\
    \boldsymbol{\tau}_{i} &= [\tau_{1,0}, \dots, \tau_{K_0,0}, \tau_{1,i}, \dots, \tau_{\triangle K_i,i}],\\
    \boldsymbol{\alpha}_{i} &= [\alpha_{1,0}, \dots, \alpha_{K_0,0}, \alpha_{1,i}, \dots, \alpha_{\triangle K_i,i}].
    \end{aligned}
    \label{features}
\end{equation}

\noindent The coefficients $\rho_{k,i}$ in the expressions $\widetilde{\gamma}_{k,0}=\rho_{k,i}\cdot\gamma_{k,0}$ model the variations of amplitude components relative to those observed in the absence of the target (see \eqref{eq:empty}). They can be used to identify and classify dead, attenuated, and unchanged components. The variables $\alpha_{q,i}$, $\tau_{q,i}$, and $\gamma_{q,i}$ represent the AoAs, ToAs/TDs, and corresponding amplitudes of the $\triangle K_{i}$ \emph{newly formed} MPCs, respectively.

In the following, we discuss an approach for the classification of birth--death, attenuated, and unchanged components.

\begin{algorithm}[tp]
    \caption{MPCs classification}
    \label{cfa} \begin{algorithmic}[1]
    \Procedure{MPC\_class}{$r(\tau,\alpha)$$,[\gamma_{k,0},\tau_{k,0},\alpha_{k,0}]_{k=1}^{K_{0}}$}
    \State $[\gamma_{q,i},\tau_{q,i},\alpha_{q,i}]_{q=1}^{K_{i}}\gets$
    CA-CFAR{[}$r(\tau,\alpha)${]}
    \For{all $K_{i}$ MPCs $\gamma_{q,i},\tau_{q,i},\alpha_{q,i}$}\Comment{Init.}
    \State \textit{MPC}$(q)$$\gets$ \textit{'unchanged'}
    \EndFor
    \For{all $K_{0}$ MPCs $\gamma_{k,0},\tau_{k,0},\alpha_{k,0}$}\Comment{Init.}
    \State \textit{dead\_MPC}$(k)$$\gets$ True
    \EndFor
    \For{all $K_{i}$ MPCs $\gamma_{q,i},\tau_{q,i},\alpha_{q,i}$}\Comment{Main}
    \For{all $K_{0}$ MPCs $\gamma_{k,0},\tau_{k,0},\alpha_{k,0}$}
    \State $\Delta \alpha \gets \alpha_{q,i} - \alpha_{k,0}$
    \State $d_{q,k} \gets \sqrt{\tau_{q,i}^{2} + \tau_{k,0}^{2} - 2\tau_{q,i}\tau_{k,0}\cos(\Delta \alpha)}$
    \If{$d_{q,k}<\varrho_{D}$} \textit{dead\_MPC}$(k)$$\gets$ False
    \EndIf
    \EndFor
    \If{$\min_{k}$$d_{q,k}<\varrho_{D}$}
    \State$\widehat{k}$$\gets\arg\min_{k}$$d_{q,k}$
    \State$\rho_{q,i}\gets$$\gamma_{q,i}/\gamma_{\widehat{k},0}|_{dB}$
    \If{$|\rho_{q,i}|>\varrho_{A}$}
    \State \textit{MPC}$(q)$$\gets$ \textit{'attenuated/enhanced'}
    \EndIf
    \Else
    \State \textit{MPC}$(q)$$\gets$ \textit{'newly\_formed}'
    \EndIf
    \EndFor
    \EndProcedure
    \end{algorithmic} 
\end{algorithm}

\subsection{MPCs Classification for Blockage and Scattering Modeling}

The proposed framework is designed to classify MPCs from noisy observations of the ADPP $r(\tau,\alpha)$ in (\ref{eq:RX}). The classifier is optimized to extract and isolate new, attenuated/enhanced, and suppressed ToA/TD and AoA features induced by a specific event $i$. We categorize the observed MPCs into four types:
\begin{enumerate}
    \item \emph{newly formed} components, 
    \item \emph{dead or suppressed} components, 
    \item \emph{attenuated or enhanced} components, 
    \item \emph{unchanged} components. 
\end{enumerate}
To discriminate between valid MPCs and noise, a denoising and multipath-search procedure is applied. In particular, the Cell-Averaging Constant False Alarm Rate (CA-CFAR) \cite{cacfar} is used to compensate for background noise, with dynamic threshold parameters summarized in Table~\ref{class} (left column).

The classification procedure follows the Algorithm~\ref{cfa}, with key parameters summarized in Table~\ref{class} (right column): 
\begin{itemize}
    \item the \emph{minimum distance} $\varrho_{D}$ between the observed MPCs
    ($\boldsymbol{\tau}_{i}, \boldsymbol{\alpha}_{i})$ and the target-free components ($\boldsymbol{\tau}_{0}$, $\boldsymbol{\alpha}_{0}$), used to evaluate whether the observed $q$-th component can
    be considered as newly formed, namely if $\min_{k}d_{q,k}>\varrho_{D}$, with
    \begin{equation}
        d_{q,k}=\sqrt{\tau_{q,i}^{2}+\tau_{k,0}^{2}-2\tau_{q,i}\tau_{k,0}\cos(\alpha_{q,i}-\alpha_{k,0})}\label{eq:distt}
    \end{equation}
    \item the \emph{amplitude tolerance} $\varrho_{A}$ used to discriminate between unchanged and attenuated (or enhanced) MPCs, as the result of the specific blockage or scattering effects. 
\end{itemize}
\textcolor{black}{All the parameter settings in Table~\ref{class} for CA-CFAR and classification were selected through a calibration phase performed on a small set of measurements acquired in the same environment used to test the static object setups (see Sect. \ref{sec:Experimental-results-in}).}

The classification algorithm (Alg.~\ref{cfa}) operates as follows. First, the $K_i$ observed components ($\boldsymbol{\gamma}_{i}, \boldsymbol{\tau}_{i}, \boldsymbol{\alpha}_{i})$ extracted using CA-CFAR are initialized as \emph{unchanged}. 
The $K_0$ reference components $(\boldsymbol{\gamma}_{0}, \boldsymbol{\tau}_{0}, \boldsymbol{\alpha}_{0})$, measured in the absence of the target, are initially flagged as \emph{dead}. Then, for each pair of components $(q,k)$, the distance $d_{q,k}$ in (\ref{eq:distt}) is computed. 
If $d_{q,k} < \varrho_D$ for some $q$, the corresponding component $k$ is no longer considered dead. Conversely, an observed component $q$ is classified as \emph{newly formed} if $\min_k d_{q,k} > \varrho_D$. Finally, amplitude variations $\rho_{q,i}$ are evaluated (in dB): if the ratio exceeds $\varrho_A$, the component is classified as \emph{attenuated} or enhanced; otherwise, it is \emph{unchanged}. The overall computational complexity of the procedure is polynomial, $\mathcal{O}(K_0 \cdot K_i)$.

\begin{figure*}
	\centering \includegraphics[scale=0.38]{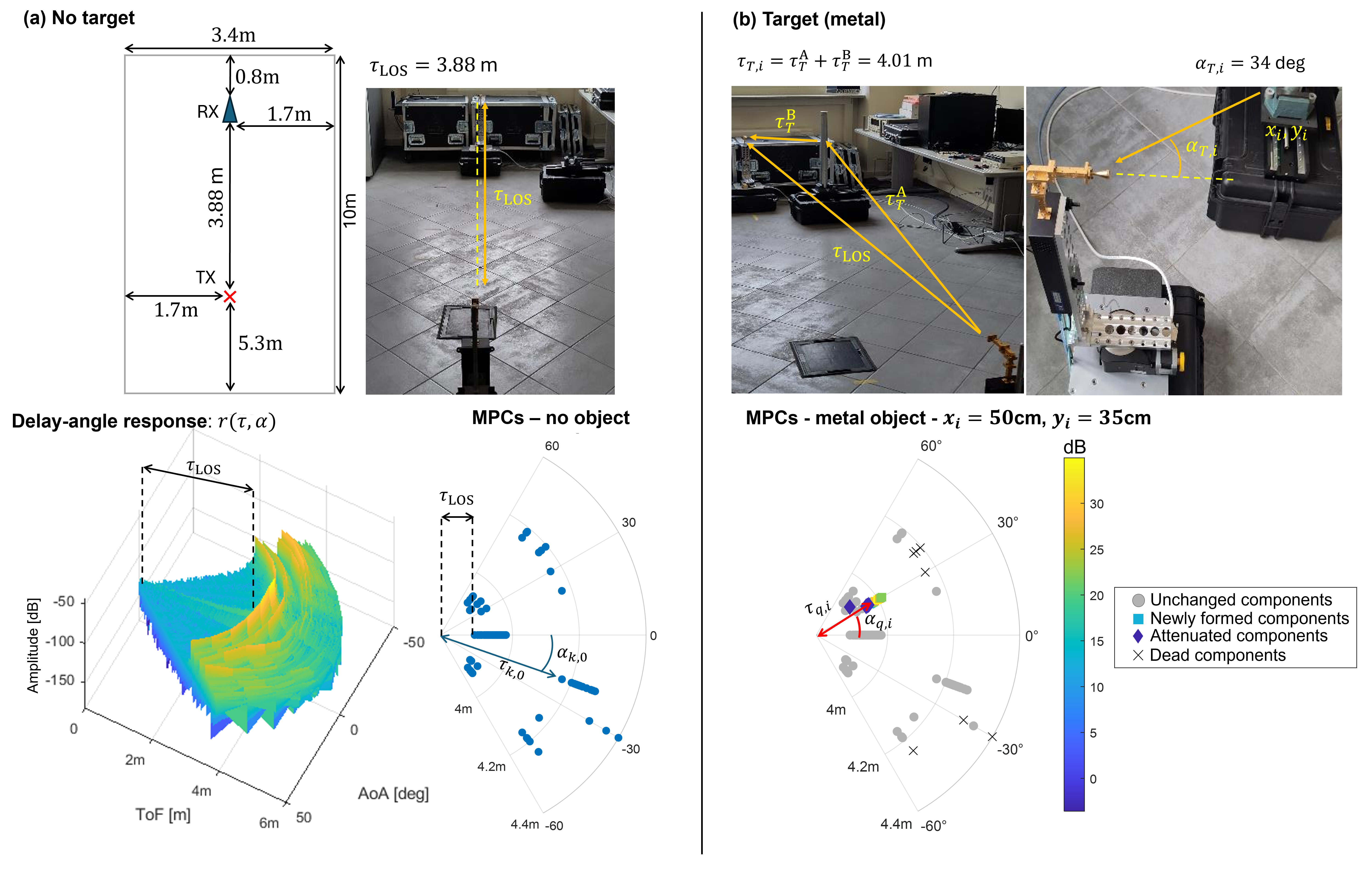} \protect\caption{\label{example-metal} \textcolor{black}{From left to right: (a) Top view: measurement setup, geometry and environment. Bottom view: ADPP response (reflectivity) [dB] $r(\tau,\alpha;t)$ and MPCs ($\tau_{k,0},\alpha_{k,0}$) for the target-free case, with LOS path length $\tau_{\mathrm{LOS}}=3.88$~m; (b) Top view: a target (metal cylinder) is located at $(x_i,y_i)=(50,35)$~cm: the object's TD is computed geometrically as $\tau_{T,i} = \tau_{T}^{\mathrm{A}} + \tau_{T}^{\mathrm{B}} = 4.01$~m, and its angle of arrival (AoA) is $\alpha_{T,i} = 34^\circ$. Bottom view: classification of MPCs $(\tau_{q,i},\alpha_{q,i})$ into unchanged, newly formed, attenuated, and dead components, with the corresponding relative amplitudes $\gamma_{q,i}$, for the considered target.} }
\end{figure*}

\begin{figure}
	\centering \includegraphics[scale=0.46]{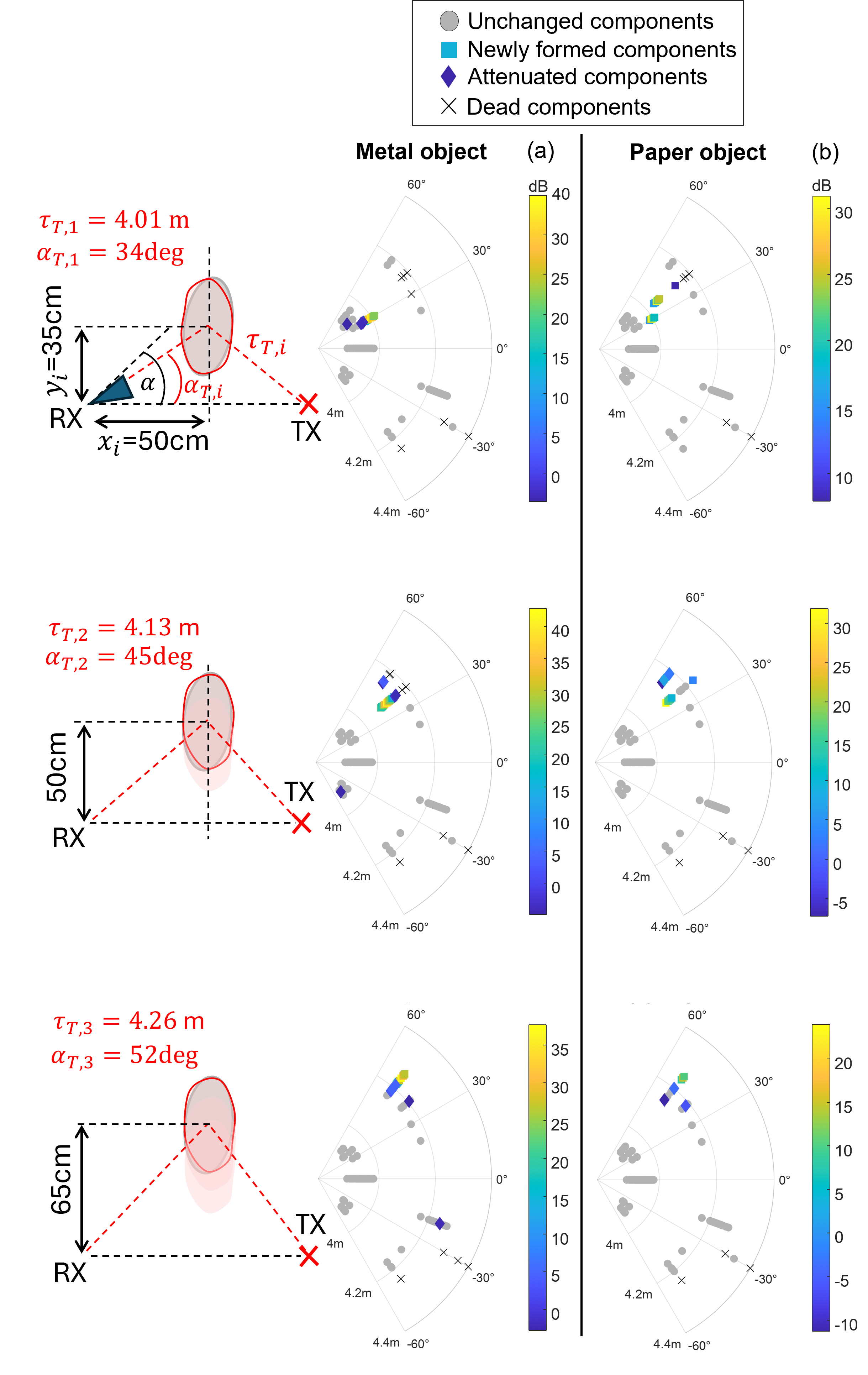} \protect\caption{\label{comp-metal-1-1} Classification of MPCs $(\tau_{q,i},\alpha_{q,i})$ for a target positioned, from top to bottom, at $(x_1,y_1) = (50,35)$~cm, $(x_2,y_2) = (50,50)$~cm, and $(x_3,y_3) = (50,65)$~cm. The corresponding object's TDs and AoAs are, respectively: $\tau_{T,1}=4.01$~m, $\alpha_{T,1}=34^\circ$; $\tau_{T,2}=4.13$~m, $\alpha_{T,2}=45^\circ$; and $\tau_{T,3}=4.26$~m, $\alpha_{T,3}=52.4^\circ$. Results are shown for metal (a) and paper (b) cylinder for comparative analysis.}
\end{figure}

\begin{figure}
	\centering \includegraphics[scale=0.46]{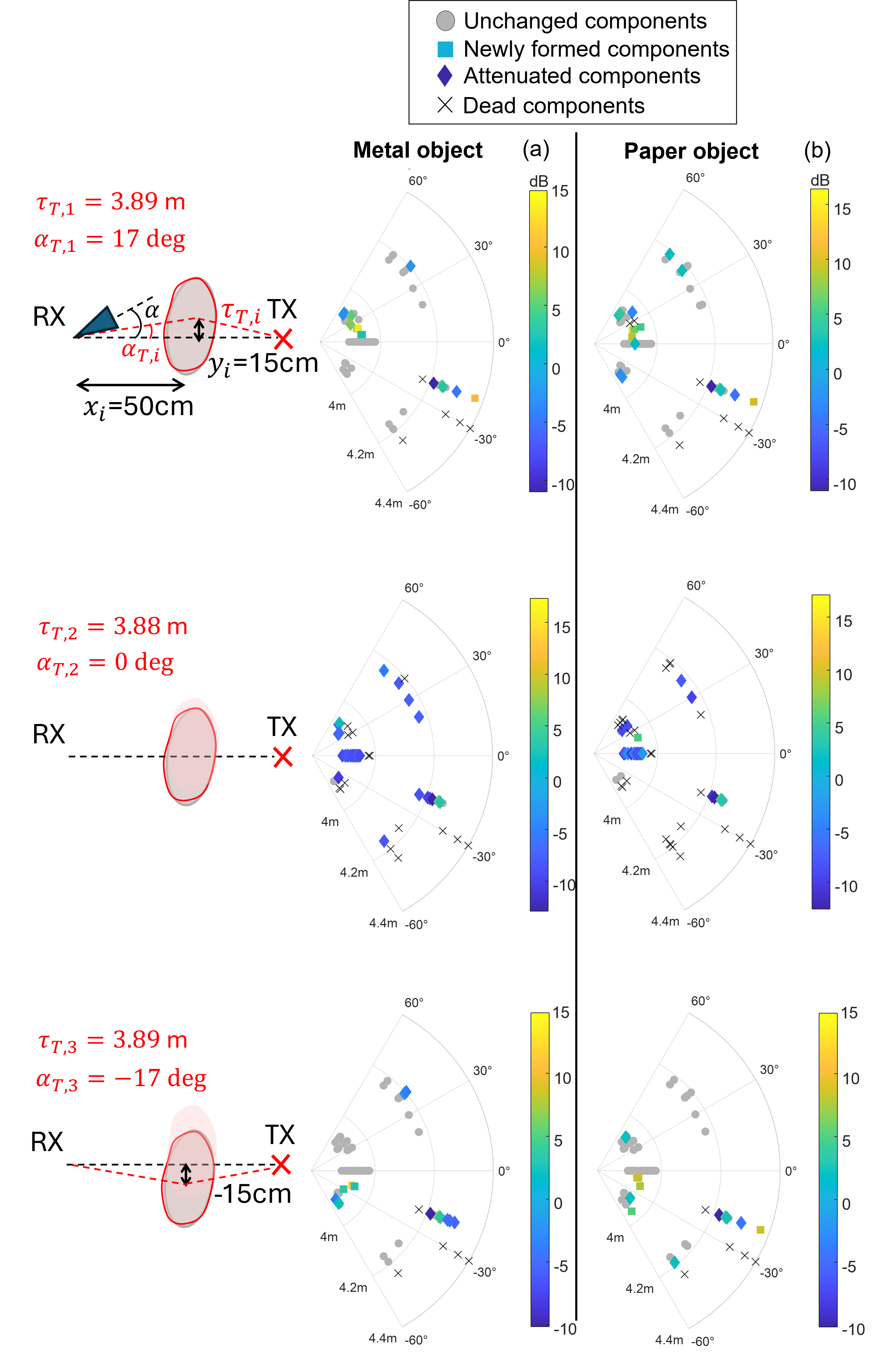} \protect\caption{\label{comp-metal-los} Classification of MPCs $(\tau_{q,i},\alpha_{q,i})$ for a target blocking the LOS path, positioned at $(x_1,y_1) = (50,15)$~cm, $(x_2,y_2) = (50,0)$~cm, and $(x_3,y_3) = (50,-15)$~cm. The corresponding object's TDs and AoAs are, respectively: $\tau_{T,1}=3.89$~m, $\alpha_{T,1}=17^\circ$; $\tau_{T,2}=\tau_{\mathrm{LOS}}$, $\alpha_{T,2}=0^\circ$; and $\tau_{T,3}=3.89$~m, $\alpha_{T,3}=-17^\circ$. Results are shown for metal (a) and paper (b) cylinder for comparative analysis.}
\end{figure}

\section{Experimental Results With Static Objects\label{sec:Experimental-results-in}}

This section discusses the results of the measurement campaign and the joint angle--delay statistics obtained for different target configurations. The analysis primarily focuses on test objects made of different materials, namely metal and paper. We address the problem of MPC classification and evaluate the positioning accuracy of the objects for different regions (ROIs), corresponding to different placements of the object with respect to the RX node. 

As introduced in Section~\ref{sec:setups}, the TX and RX nodes are separated by a line-of-sight (LOS) distance of $\tau_{\mathrm{LOS}}=3.88$~m. The actual position of the blocker/object is expressed in terms of its relative coordinates $(x_{i},y_{i})$, as well as the corresponding AoA $\alpha_{T,i}$ and TD $\tau_{T,i}$ between TX and RX. The TD is defined as the geometric path length from TX to RX intercepting the object at its barycenter $(x_{i},y_{i})$. The pair $(\alpha_{T,i},\tau_{T,i})$ thus identifies the most significant component introduced by the blocker. 

The adopted approach for target localization consists of reconstructing the features $(\alpha_{T,i},\tau_{T,i})$ from the classified MPCs obtained via Algorithm~\ref{cfa} and subsequently estimating the blocker position $(\widehat{x}_{i},\widehat{y}_{i})$. It is worth noting that the MPCs extracted from the noisy channel response $r(\tau,\alpha)$ inherently depend on both the state of the blocker and on its interaction with the surrounding environment.

In the following section, we extend the study to the case of a human body and investigate the impact of unintentional movements on the joint angle--delay statistics. 

\subsection{Multipath Component Classification \label{sec:MPC_class}}

An example of MPC classification is shown in Fig.~\ref{example-metal}. Fig.~\ref{example-metal}(a) shows the noisy angle--delay response $r(\tau,\alpha)$ in the absence of the target, together with the corresponding MPCs $(\boldsymbol{\tau}_{0},\boldsymbol{\alpha}_{0})$ obtained by CA-CFAR using the parameters listed in Table~\ref{class}. The LOS path is clearly visible at $\alpha=0$ and TD = $\tau_{\textrm{LOS}}$. The contributions exceeding $\tau_{\textrm{LOS}}$ are caused by reflections off the metallic parts of the measurement supports. 
Moreover, it is apparent that several scattered contributions due to both furniture and walls are also present in the measurement scene up to $4.4$ m and at $|\alpha|>20^\circ$.

Fig.~\ref{example-metal}(b) considers instead a cylindrical metal target located at $(x_{i},y_{i})=(50,35)$~cm. Newly formed (square markers), dead (cross markers), attenuated/enhanced (diamond markers), or unchanged (gray markers) MPCs are classified according to Algorithm~\ref{cfa} with parameters in Table~\ref{class}. For clarity, the classified MPCs $(\tau_{q,i},\alpha_{q,i})$ are shown in terms of TDs and AoAs over the intervals 3.8--4.4~m and $[-60^{\circ},\,60^{\circ}]$, respectively. 
The amplitudes $\gamma_{q,i}$ are defined with respect to the target-free case: newly formed or enhanced components have positive gains ($\gamma_{q,i}>0$~dB), while attenuated components have negative gains ($\gamma_{q,i}<0$~dB). These effects are visually distinguished using distinct colors.
For the considered event $i$, the TD and AoA of the object correspond to $\tau_{T,i}= \tau_{T}^{\mathrm{A}}+\tau_{T}^{\mathrm{B}}=4.01$~m and $\alpha_{T,i}=34^{\circ}$ which is consistent with the set position of the linear translation stage.

Figures~\ref{comp-metal-1-1} and \ref{comp-metal-los} summarize the classified MPCs for metal and paper cylinders positioned 50~cm from the RX, with different offsets relative to the LOS path. 
\\Specifically, offsets are $y_{1}=35$~cm, $y_{2}=50$~cm, and $y_{3}=65$~cm (Fig.~\ref{comp-metal-1-1}), and $y_{4}=-15$~cm, $y_{5}=0$~cm, and $y_{6}=15$~cm (Fig.~\ref{comp-metal-los}). 
The metal and paper cylinders have diameters and heights of $5\times72$~cm and $5.5\times91$~cm, respectively. 
The results across these cases reveal the following: 
\begin{enumerate}
\item \textcolor{black}{As depicted in Fig.~\ref{comp-metal-1-1}, scatterer positions with
offsets $\lvert y\rvert\geq15$~cm from the LOS can be determined
from the newly formed MPCs, as these directly reflect the target geometric
TD. 
Target micro-movements
as small as 50~mm can be also detected, with a theoretical range-resolution
limit of 2~mm for the considered $B=70$~GHz band. The impact of
different operating bandwidths is analyzed in the next section in
the context of passive body localization.} 
\item \textcolor{black}{Metal objects generate newly formed MPCs with significantly larger
amplitudes than paper objects as their higher conductivity produces
stronger reflections. Table~\ref{metal-vs-paper_tab} reports the average amplitudes of the
newly formed components $\overline{\gamma}_{q,i}=\bigtriangleup K_{i}^{-1}\sum_{q}\gamma_{q,i}$
due to metal and paper scatterers, respectively: when the target is
close to the LOS, the amplitude of these components can be up to $10$
dB higher for metal than for paper, while the difference becomes less
pronounced ($2$dB or less) as the object moves farther away from
the LOS. This highlights the capability to discriminate between materials
with substantially different EM properties (metal vs. paper), although
this aspect is not further explored in this work.} 
\item \textcolor{black}{As shown in Fig.~\ref{comp-metal-los}, targets fully or partially
obstructing the LOS ($\lvert y\rvert<15$~cm), i.e., blocking the
first Fresnel region in the considered band, cause the disappearance
of LOS components ($\alpha_{q,i}=0^{\circ}$) and/or the attenuation
of MPCs, whose gain variations $\gamma_{q,i}$ can be interpreted
using diffraction-based considerations \cite{ramp}. Material discrimination
for objects blocking the LOS within the Fresnel region should primarily
rely on the analysis of attenuated and suppressed MPC rather than
on newly formed ones since material-dependent absorption effects may
dominate.}
\end{enumerate}

\subsection{Positioning Accuracy Analysis\label{subsec:pos_acc}}

In this section, we address the problem of inferring the relative position
$(\widehat{x}_{i},\widehat{y}_{i})$ of an object located outside the Fresnel region.\footnote{Detecting objects within the Fresnel region primarily relies on monitoring dead and attenuated MPCs and basic diffraction principles; hence, this case is not considered here.} The estimation is based on the classified MPCs from (\ref{features}). 

First, we estimate the object TD $\widehat{\tau}_{T,i}$ and angle $\widehat{\alpha}_{T,i}$ from the response $r(\tau,\alpha;t)$. Among the $\triangle K_{i}$ newly formed MPCs, characterized by attenuations $\boldsymbol{\gamma}_{\triangle K_{i}}=\left[\gamma_{1,i},\ldots,\gamma_{\triangle K_{i},i}\right]$, ToA/TD $\boldsymbol{\tau}_{\triangle K_{i}}=\left[\tau_{1,i},\ldots,\tau_{\triangle K_{i},i}\right]$, and AoAs $\boldsymbol{\alpha}_{\triangle K_{i}}=\left[\alpha_{1,i},\ldots,\alpha_{\triangle K_{i},i}\right]$, the pair $(\widehat{\tau}_{T,i},\widehat{\alpha}_{T,i})$ is defined by the component with the maximum amplitude:
\begin{equation}
    \widehat{\tau}_{T,i}=\tau_{\widehat{q},i},\quad \widehat{\alpha}_{T,i}=\alpha_{\widehat{q},i} \quad \text{with} \quad \widehat{q}=\arg\max_{q\in[1,\ldots,\triangle K_{i}]}\gamma_{q,i}.
    \label{eq:maxtd}
\end{equation}

Next, the relative target position is extracted by solving:
\begin{equation}
    \left\{
    \begin{aligned}
    & \sqrt{(\widehat{x}_{i}-\tau_{\mathrm{LOS}})^{2}+\widehat{y}_{i}^{2}}
       + \rho_{\widehat{x}_{i},\widehat{y}_{i}}
       = \widehat{\tau}_{T,i}, \\[1ex]
    & (\widehat{x}_{i}-\tau_{\mathrm{LOS}})^{2} + \widehat{y}_{i}^{2}
       = \tau_{\mathrm{LOS}}^{2} + \rho_{\widehat{x}_{i},\widehat{y}_{i}}^{2} -\\ 
    & \quad {} - 2\,\tau_{\mathrm{LOS}}\,
       \rho_{\widehat{x}_{i},\widehat{y}_{i}}
       \cos(\widehat{\alpha}_{T,i}) .
    \end{aligned}
    \right.
    \label{eq:trig}
\end{equation}
for the unknowns $(\widehat{x}_{i},\widehat{y}_{i})$, where 
$\rho_{\widehat{x}_{i},\widehat{y}_{i}}=\sqrt{\widehat{x}_{i}^{2}+\widehat{y}_{i}^{2}}$, under the constraint that the object azimuth angle lies within $[-60^{\circ},\,60^{\circ}]$.

\begin{figure}
	\centering
    \includegraphics[scale=0.45]{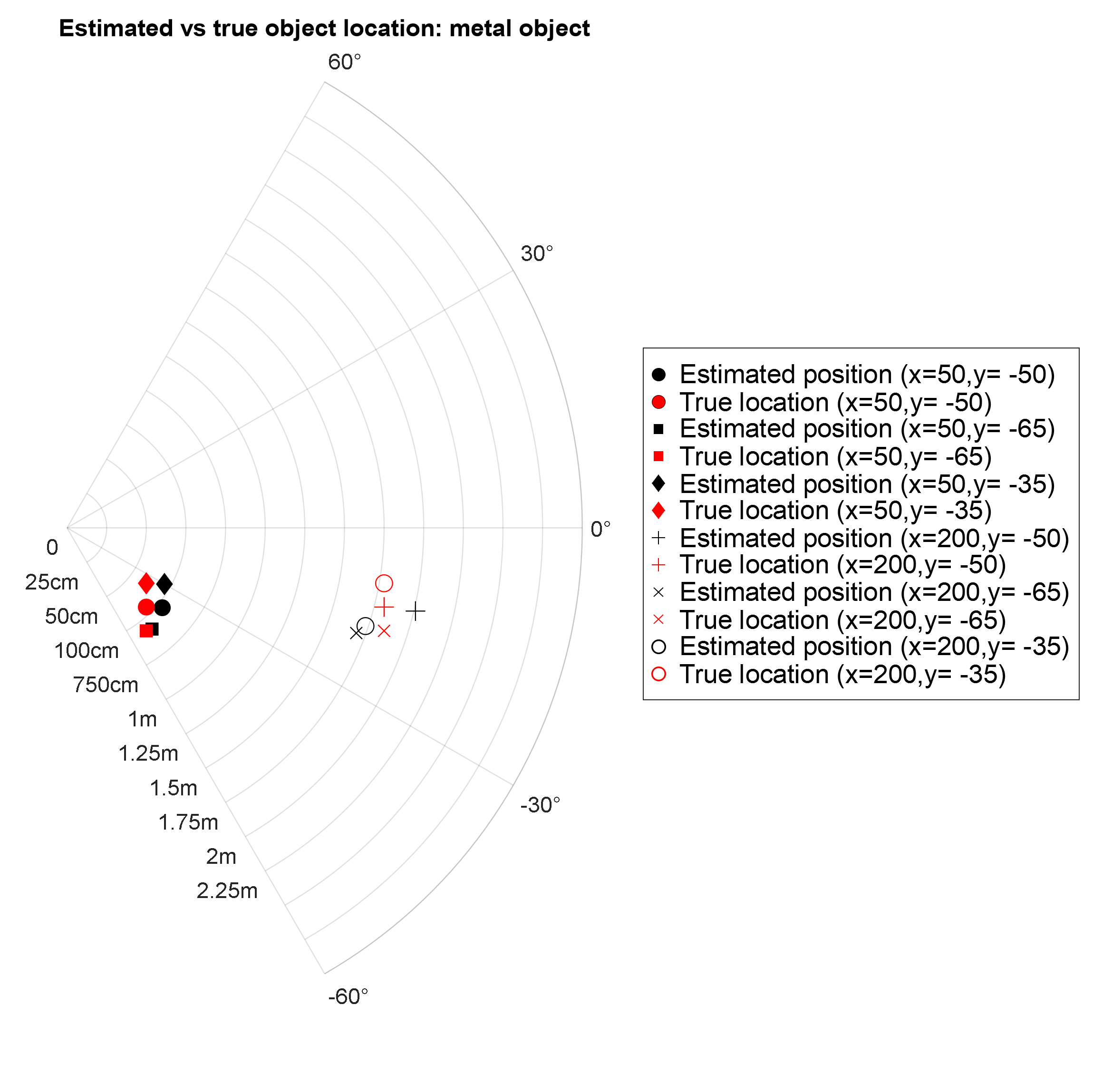} \protect\caption{\label{loc_ex} Example of object localization for six target positions. Red markers indicate the true object locations: $(x_{1},y_{1})=(50,-35)$~cm (diamonds), $(x_{2},y_{2})=(50,-50)$~cm (circles), $(x_{3},y_{3})=(50,-65)$~cm (squares), $(x_{4},y_{4})=(200,-35)$~cm (empty circles), $(x_{5},y_{5})=(200,-50)$~cm (plus signs), and $(x_{6},y_{6})=(200,-65)$~cm (crosses). Black markers show the corresponding estimated positions from (\ref{eq:trig}). A metallic object is considered.}
\end{figure}

Fig.~\ref{loc_ex} illustrates an example of object localization for $6$ positions: black markers denote estimated positions, while red markers indicate the true locations of the metallic object. 
\\Positioning accuracy, expressed in terms of Root Mean Square Error (RMSE), is reported in Table~\ref{pos_acc} for both metallic and paper objects. For objects close to the RX ($x_{i=1,2,3}=50$~cm), the average RMSE is 8.4~cm for metal and 21.5~cm for paper. Errors increase when the object is farther from the RX, especially when equidistant from TX and RX (i.e., $x_{i=4,5,6}=200$~cm), yielding RMSEs of 19.6~cm (metal) and 33.7~cm (paper). Overall, the average RMSE across the considered ROIs is 16.3~cm for metal and 28.6~cm for paper.

\begin{table}[tp]
\caption{\textcolor{black}{Average newly formed MPCs gain $\overline{\gamma}_{q,i}$ {[}dB{]}
variation analysis for metal and paper objects.}}\label{metal-vs-paper_tab}
\centering %
\begin{tabular}{lcc}
\hline\hline
\textbf{True location (cm)} & \textbf{Metal} & \textbf{Paper}\tabularnewline
 & \textbf{$\overline{\gamma}_{q,i}$ (dB)} & \textbf{$\overline{\gamma}_{q,i}$ (dB)}\tabularnewline
\hline 
$(x,y)=(50,-50)$ & 26.3 & 16.4\tabularnewline
$(x,y)=(50,-65)$ & 23.6 & 21.3\tabularnewline
$(x,y)=(50,-35)$ & 25.4 & 14.6\tabularnewline
$(x,y)=(200,-50)$ & 10.7 & 11.1\tabularnewline
$(x,y)=(200,-65)$ & 7.7 & 9.4\tabularnewline
$(x,y)=(200,-35)$ & 14.1 & 10.6\tabularnewline
\hline\hline
\end{tabular}
\end{table}

\begin{table}[tp]
    \caption{\label{pos_acc}Positioning accuracy in terms of RMSE for metal and paper cylinders at six nominal locations}
    \centering
    \begin{tabular}{lcc}
    \hline\hline
    \textbf{True location (cm)} & \textbf{RMSE, metal} & \textbf{RMSE, paper} \\
    \textbf{} & \textbf{(cm)} & \textbf{(cm)} \\
    \hline
    $(x,y)=(50,-50)$  & 10 & 10.1 \\
    $(x,y)=(50,-65)$  &  3.8 & 24.5 \\
    $(x,y)=(50,-35)$  & 11.4 & 30 \\
    $(x,y)=(200,-50)$ & 19.9 & 58.9 \\
    $(x,y)=(200,-65)$ & 17.5 & 25.9 \\
    $(x,y)=(200,-35)$ & 29.4 & 20.1 \\
    \hline\hline
    \end{tabular}
\end{table}

\section{Applications to Body Sensing\label{sec:bodysensing}}

In this section we consider the problem of passive body localization. Compared to static object detection, voluntary and involuntary body motions introduce a larger number of new and dead MPCs, necessitating an angle--delay statistical characterization. The true nominal body location is again reported in terms of the relative coordinates ($x_{i},y_{i}$); however, we also account for the (non-negligible) lateral and anteroposterior dimensions of the body, approximated as $20$~cm and $12.5$~cm, respectively \cite{ramp}. 
\\For this purpose, we denote $t$ as the time at which a specific ADPP response is acquired by the VNA. In the following, we analyze the classified MPCs extracted from the noisy response $r(\tau,\alpha;t)$ observed over consecutive time instants $t$.

\subsection{Motion Effects on Birth--Death of Multipath Components}

\begin{figure}[t]
	\centering 
    \includegraphics[scale=0.45]{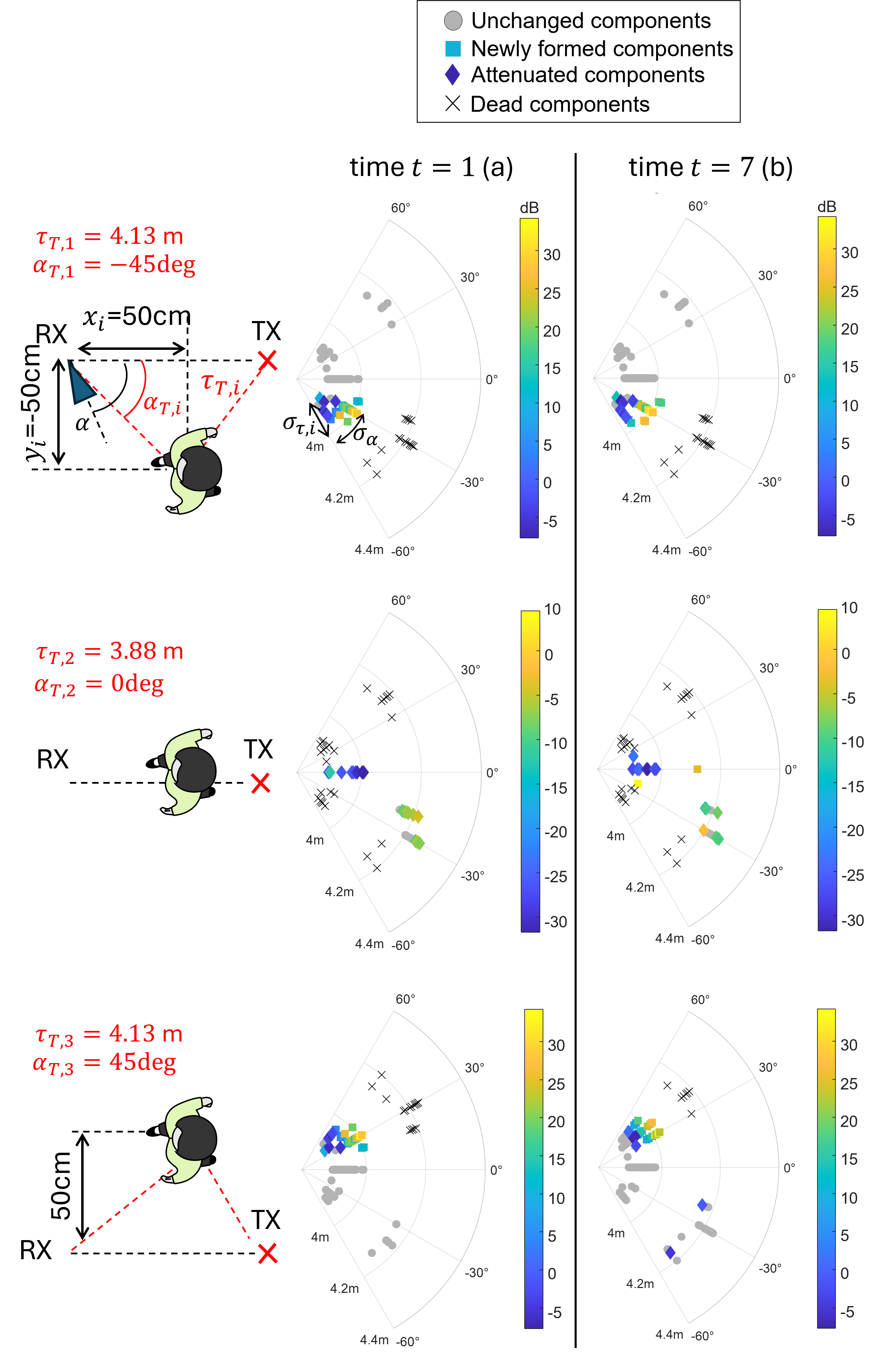} \protect\caption{\label{comp-body1} Classification of MPCs $(\tau_{q,i},\alpha_{q,i})$ for a body subject located at $(x_{1},y_{1})=(50,-50)$~cm, $(x_{2},y_{2})=(50,0)$~cm, and $(x_{3},y_{3})=(50,50)$~cm (from top to bottom). The subject performs small movements around each nominal position. Results at two time instants, $t=1$~s (a) and $t=7$~s (b), are shown for comparison.}
\end{figure}

\begin{figure}[t]
	\centering
    \includegraphics[scale=0.45]{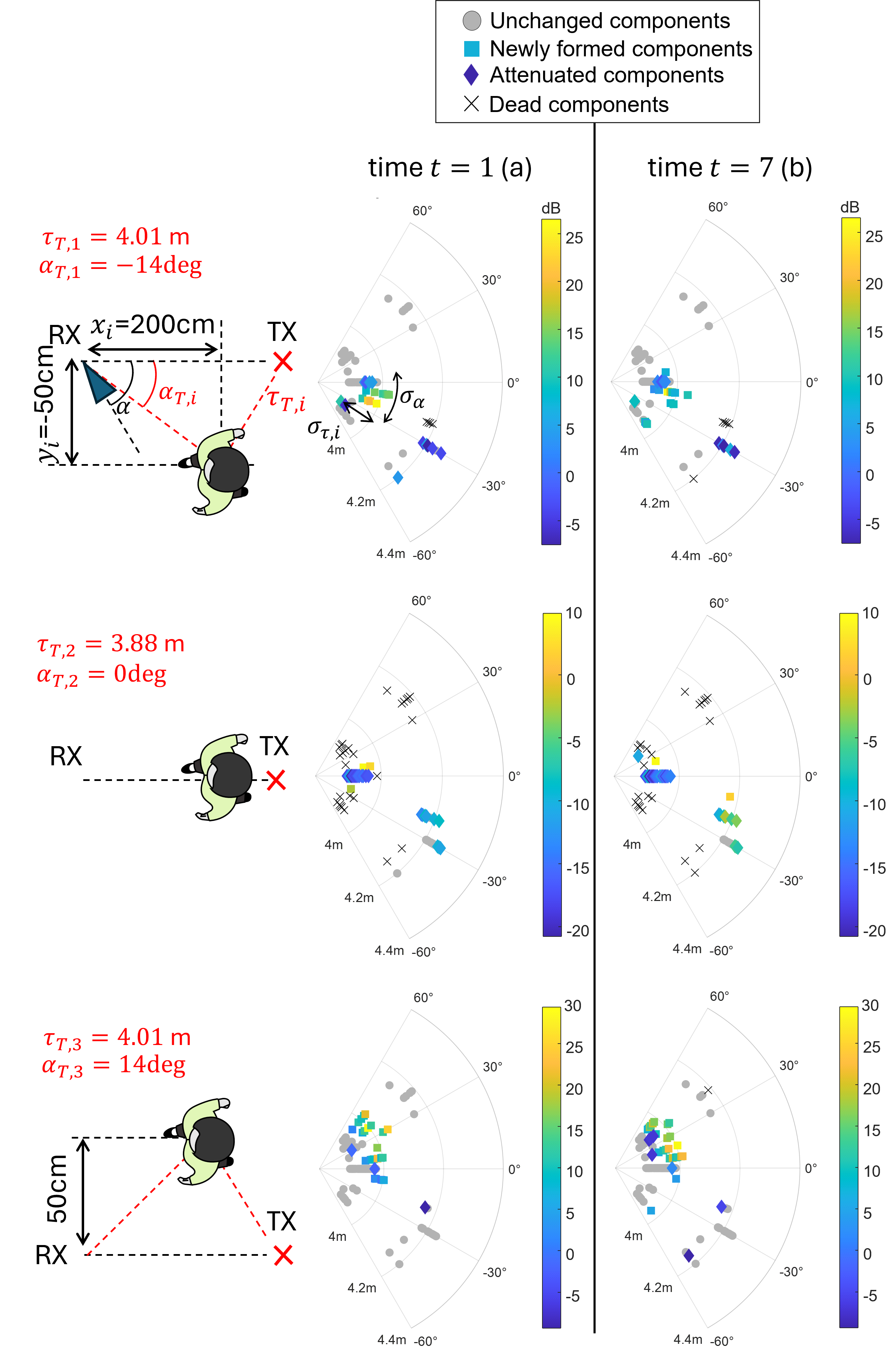} \protect\caption{\label{comp-body2} Classification of MPCs $(\tau_{q,i},\alpha_{q,i})$ for a body subject located at $(x_{1},y_{1})=(200,-50)$~cm, $(x_{2},y_{2})=(200,0)$~cm, and $(x_{3},y_{3})=(200,50)$~cm (from top to bottom). The subject performs small movements around each nominal position. Results at two time instants, $t=1$~s (a) and $t=7$~s (b), are shown for comparison.}
\end{figure}

Figures~\ref{comp-body1} and \ref{comp-body2} summarize the classified MPCs for a human subject located at $x_{i}=50$~cm (Fig.~\ref{comp-body1}) and $x_{i}=200$~cm (Fig.~\ref{comp-body2}) from the RX, with offsets $y_{i}=0$~cm and $y_{i}=50$~cm relative to the LOS path. The impact of body movements around the nominal positions can be observed through two distinct RF measurements taken at a $10$-second interval ($t=1$ and $t=7$, from left to right). Based on the object detection analysis, the following conclusions can be drawn:

\begin{enumerate}
    \item A subject obstructing the LOS, or with offsets $\left|y\right|<15$~cm from the LOS, results in dead LOS components, similar to a perfectly absorbing blockage, and can be modeled via diffraction effects. The ToA/TD of the \emph{dead} components, $\tau_{q,i} \geq \tau_{\mathrm{LOS}}$, can be modeled as exponentially distributed:
    \begin{equation}
        \tau_{q,i}-\tau_{\mathrm{LOS}}\sim\mathrm{\varSigma}(\cdot|\sigma_{\tau}^{(d)})
        = \frac{1}{\sigma_{\tau}^{(d)}}\exp\Big(-\frac{x}{\sigma_{\tau}^{(d)}}\Big),
        \label{eq:dead}
    \end{equation}
    with deviation $\sigma_{\tau}^{(d)}$ reported in Table~\ref{deviations}. Note that exponential distribution is typically considered for Power Delay Profile (PDP) modeling of cluster delays \cite{etsi}. The exponential distribution is here applied to model the dead MPCs induced by objects obstructing the LOS.
    
    \item Newly formed MPCs are observed for body positions with offsets $\left|y\right|\geq15$~cm from the LOS. In addition, involuntary body movements generate several new MPCs whose ToA/TD and AoA features can be modeled 
    as Gamma and Laplace distributed, respectively:
    \begin{equation}
        \tau_{q,i}-\tau_{\mathrm{LOS}}\sim\Gamma(\cdot|\tau_{T,i},\sigma_{\tau,i}^{(b)}),
        \label{eq:gamma}
    \end{equation}
    with $\tau_{T,i}$ and $\sigma_{\tau,i}^{(b)}$ denoting the mode and standard deviation of the Gamma distribution \cite{chenstat}, and
    \begin{equation}
        \alpha_{q,i}
        \sim \mathrm{L}(\cdot|\alpha_{T,i},\sigma_{\alpha}^{(b)})=\frac{1}{2\sigma_{\alpha}^{(b)}}\exp\Big(-\frac{|x-\alpha_{T,i}|}{\sigma_{\alpha}^{(b)}}\Big),
        \label{eq:beta}
    \end{equation}
    with $\alpha_{T,i},\sigma_{\alpha}^{(b)}$ indicating the corresponding mean and standard deviation of the Laplace distribution. Gamma, Laplace functions and their variants are typically adopted to model shadowing effects (Nakagami-m fading) and angular/azimuth spreads \cite{laplace}. As for object detection, $\tau_{T,i}$ and $\alpha_{T,i}$ correspond to the geometric TD and AoA with respect to the body barycenter at $(x_{i},y_{i})$, and can be used to estimate the true relative body position as in (\ref{eq:trig}).
\end{enumerate}
\textcolor{black}{The adopted exponential \eqref{eq:dead}, Gamma \eqref{eq:gamma}, and Laplace \eqref{eq:beta} distributions are inspired by well-established statistical channel models, including 3GPP frameworks for delay and angular spreads \cite{etsi}. On the other hand, their selection is also supported by empirical data. In what follows, the distributions are fitted to the measured angle-delay MPCs data.}

Table~\ref{deviations} reports the observed deviations for dead ($\sigma_{\tau}^{(d)}$) and newly formed ($\sigma_{\tau,i}^{(b)}$, $\sigma_{\alpha}^{(b)}$) MPCs estimated from measurements. Note that ToA/TD deviations of new components depend on the distance $x_{i}$ of the body from the RX, ranging from 1.5~cm (for a subject equidistant from TX and RX, $x_{i}=2$~m) to 5.8~cm (for a body close to RX, $x_{i}=0.5$~m). The corresponding AoA deviations are instead less affected, with $\sigma_{\alpha}^{(b)} \in [10,14]^\circ$.

\begin{figure}
	\centering \includegraphics[scale=0.35]{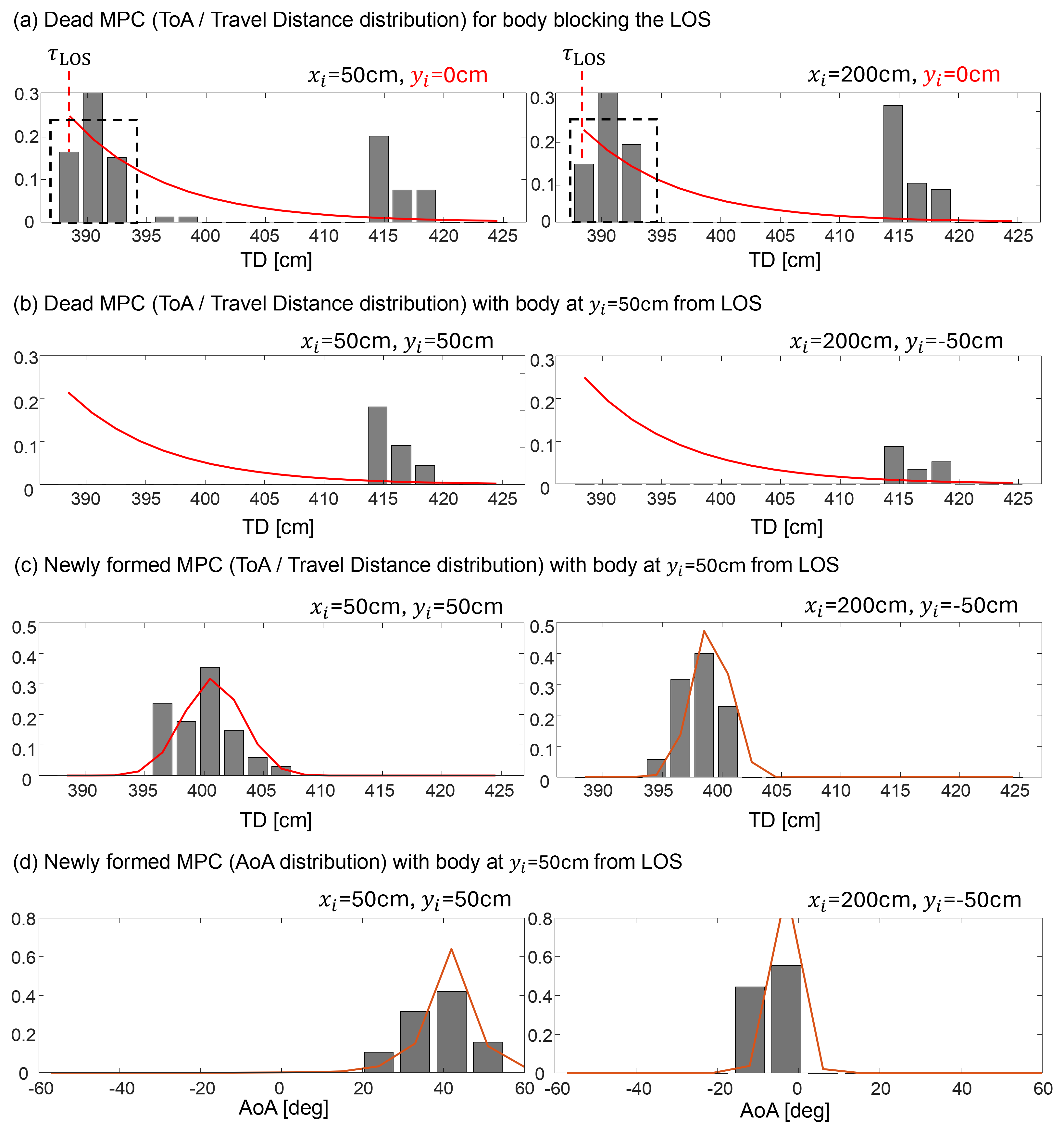} \protect\caption{\label{comp-stat} Statistical characterization of dead and newly formed MPCs. Empirical probability mass function from measurements (gray bars) compared with exponential (\ref{eq:dead}), Gamma (\ref{eq:gamma}) and Laplace (\ref{eq:beta}) distributions (red lines). From top to bottom: (a) dead MPC TD distribution for body obstructing the LOS path ($y_{i}=0$~cm) at distance $x_{i}=50$~cm (left) and $x_{i}=200$~cm (right) from the RX. Exponential distribution model fit for the near-LOS paths (dashed box) is superimposed. (b) Dead MPC TD distribution for body with offset $y_{i}=50$~cm from LOS and same distances from the RX. Note that the model does not fit, as LOS components are unaffected according to (\ref{eq:dead}). (c) Newly formed MPC TD distribution for body at same positions as in (b) and Gamma model fit. (d) Newly formed MPC AoA distribution for body at same positions and Laplace model fit.}
\end{figure}

The statistical characterization of newly formed and dead MPCs is further analyzed in Fig.~\ref{comp-stat}. Fig.~\ref{comp-stat}(a) shows the empirical distribution of ToAs/TDs $\tau_{q,i}$ of MPCs classified as suppressed when the body blocks the LOS ($y_{i}=0$) at $x_{i}=50$~cm and $x_{i}=200$~cm. Two main clusters of suppressed components are observed: the first, highlighted with dashed lines, corresponds to near-LOS paths ($\tau_{\mathrm{LOS}}=38.8$~cm), while the second cluster arises from components reflected from the side wall ($\tau=412$~cm). The exponential distributions (\ref{eq:dead}), shown in red lines, align with the first cluster, consistent with their role in modeling LOS blockage effects. 

When the body is positioned outside the Fresnel region ($y_{i}=50$~cm), the LOS components are either unaffected or attenuated, but no longer categorized as suppressed. Fig.~\ref{comp-stat}(b) shows the empirical distribution of TDs and AoAs of new MPCs for two body positions. Gamma (\ref{eq:gamma}) and Laplace (\ref{eq:beta}) distributions (red lines) with parameters from Table~\ref{deviations} are chosen to model the TDs/ToAs and AoAs, respectively, and are superimposed for each case.

\begin{table*}[tp]
    \caption{\label{deviations} TD and AoA deviations of dead and new MPCs compared with true target locations}
    \centering
    \begin{tabular}{l l c c c c}
    \hline\hline
     & \textbf{True location (cm)} & $\tau_{T,i}$ (cm) & $\alpha_{T,i}$ (degs) & $\sigma_{\tau,i}$ (cm) & $\sigma_{\alpha}$ (degs) \\
    \hline
    \textbf{Dead MPCs} 
     & $(x,y)=(50,0)$ & n.a. & n.a. & $\sigma_{\tau,i}^{(d)}=1.1$ & n.a. \\
     & $(x,y)=(200,0)$ & n.a. & n.a. & $\sigma_{\tau,i}^{(d)}=1.1$ & n.a. \\
    \textbf{New MPCs} 
     & $(x,y)=(50,50)$ & $401.2$ & $25.1$ & $\sigma_{\tau,i}^{(b)}=2.7$ & $\sigma_{\alpha}^{(b)}=14.3$ \\
     & $(x,y)=(50,-50)$ & $400.7$ & $-29.8$ & $\sigma_{\tau,i}^{(b)}=2.5$ & $\sigma_{\alpha}^{(b)}=13.7$ \\
     & $(x,y)=(200,50)$ & $399$ & $11.4$ & $\sigma_{\tau,i}^{(b)}=1.6$ & $\sigma_{\alpha}^{(b)}=10.7$ \\
     & $(x,y)=(200,-50)$ & $397.7$ & $-12$ & $\sigma_{\tau,i}^{(b)}=1.2$ & $\sigma_{\alpha}^{(b)}=10.3$ \\
    \hline\hline
    \end{tabular}
\end{table*}

\begin{figure}
	\centering
    \includegraphics[scale=0.55]{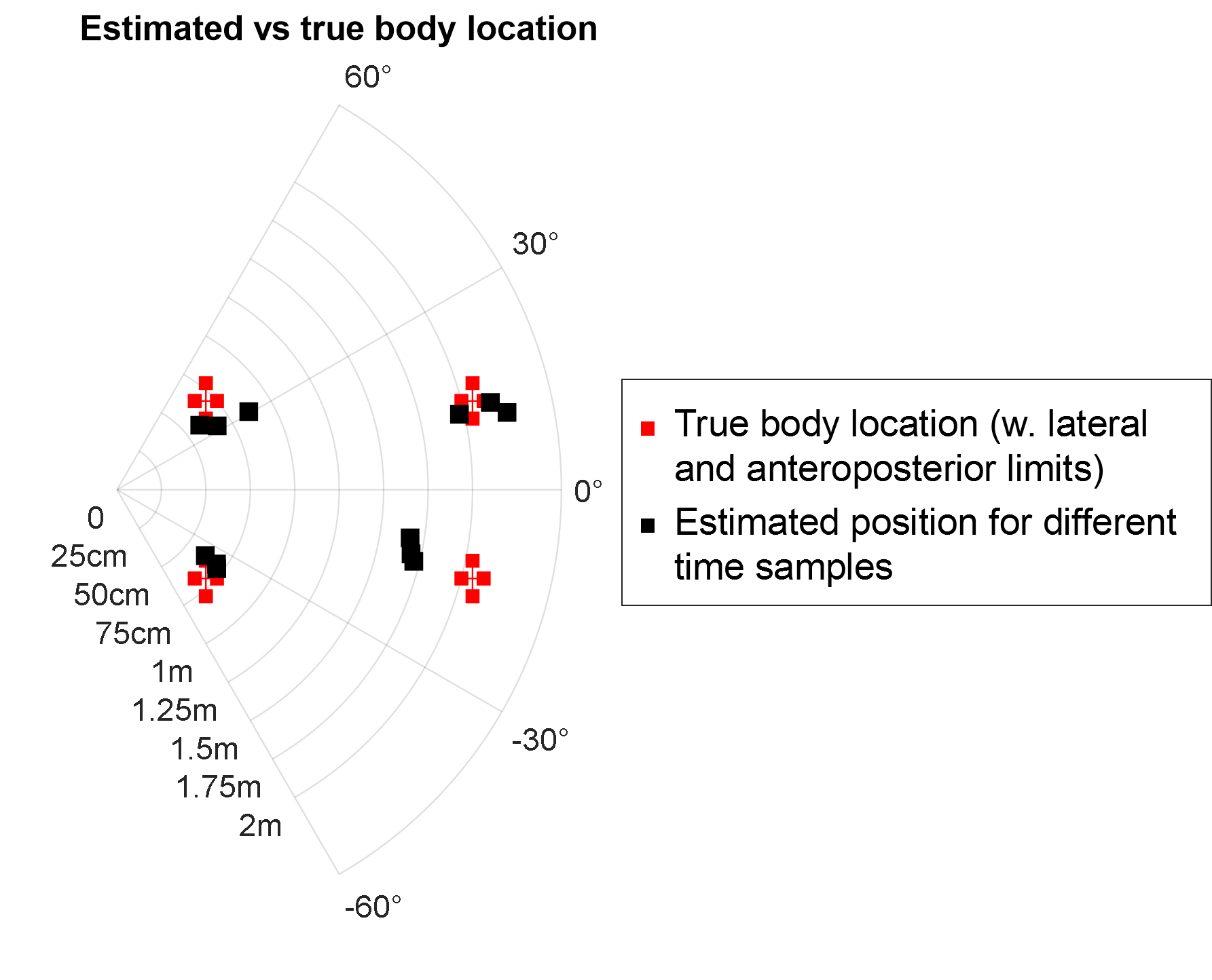}
	\protect\caption{\label{localization} Example of body localization for four subject positions. Red markers indicate the area occupied by the subject, approximated with lateral and anteroposterior dimensions of $20$~cm. Nominal positions are $(x,y)=(50,-50)$~cm, $(50,50)$~cm, $(200,-50)$~cm, and $(200,50)$~cm. Black markers show the corresponding estimated positions obtained with (\ref{eq:trig}) over consecutive RF acquisitions.}
\end{figure}
\subsection{Positioning Accuracy and Operating Bandwidth}\label{subsec:posacc}

The approach we follow for body localization consists of three stages: 
\begin{enumerate}
	\item Extract the newly formed MPCs from the measured ADPP response $r(\tau,\alpha;t)$ at time $t$ using Algorithm~\ref{cfa}; 
 	\item Implement a Log-likelihood Ratio Test (LRT) 
    using the $S$ suppressed MPCs $\tau_{q,i}$ to discriminate whether the body is blocking the LOS path ($\left|y\right|<15$~cm) and causing diffraction effects, or moving outside the Fresnel zone of the link. Using (\ref{eq:dead}) the LRT becomes:
	\begin{equation}
	\left\{ \begin{array}{ll}
	\hspace{-0.4em}\sum_{q}\left(\tau_{q,i}-\tau_{\mathrm{LOS}}\right)\leq S\kappa & \quad\textrm{blocking the LOS}\\
	\hspace{-0.4em}\sum_{q}\left(\tau_{q,i}-\tau_{\mathrm{LOS}}\right)>S\kappa & \quad\textrm{outside Fresnel area}
	\end{array}\right.\label{eq:los-nlos}
	\end{equation} 
	with $\kappa \approx 30 \sigma_{\tau}^{(d)}$~mm from measurements;
	\item Reconstruct the TD $\widehat{\tau}_{T,i}$ and the AoA $\widehat{\alpha}_{T,i}$ of the body according to (\ref{eq:maxtd}) and then obtain the subject relative position $(\widehat{x}_{i},\widehat{y}_{i})$ by solving (\ref{eq:trig}).
\end{enumerate}

Fig.~\ref{localization} shows an example of body localization for the positions listed in Table~\ref{deviations}. Black markers indicate the estimated locations at different time instants, while red markers indicate the boundary of the body accounting for lateral and anteroposterior dimensions. The average positioning RMSE for a body close to the RX ($x_{i}=50$~cm) is $12.1$~cm. As observed for object detection, the error increases for a subject equidistant from both the TX and RX (i.e., $x_{i}=200$~cm) to $26.6$~cm. \textcolor{black}{Although the extension to continuous-time tracking and explicit reconstruction of the blocker trajectory is not addressed in this work, we observe that consecutive MPC observations could be exploited by incorporating a prior motion model for the human body, enabling trajectory smoothing and outlier rejection in the reconstructed trajectory positions.}

\begin{figure}
    \centering
    \includegraphics[scale=0.60]{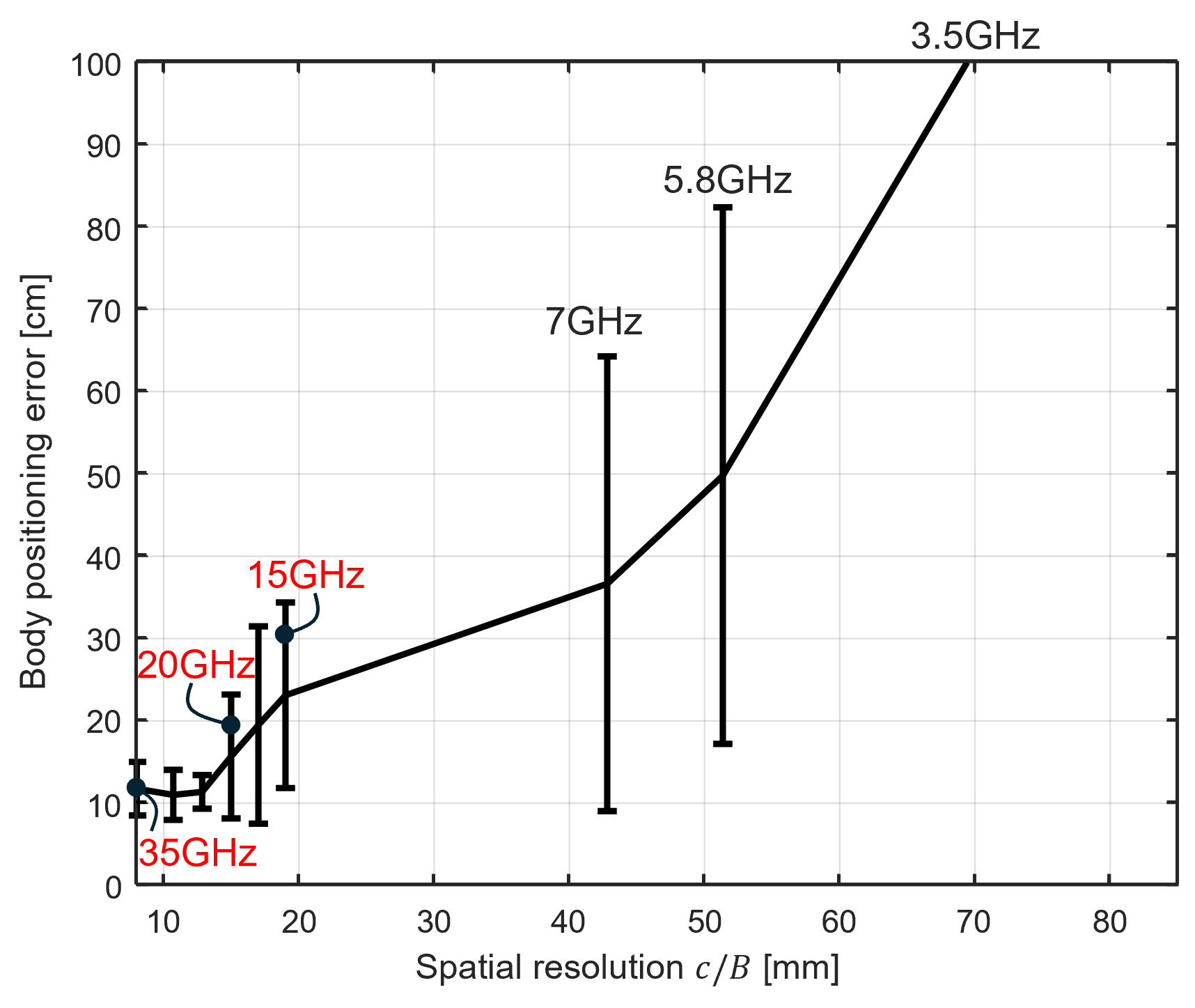} \protect\caption{\label{range_resol}\textcolor{black}{Average positioning error with error bars (cm) as a function of the corresponding spatial resolution $c/B$ (mm) in free space, with $c$ being the light speed and $B$ the operating bandwidth.}}
\end{figure}

In Fig.~\ref{range_resol} we analyze the average positioning error as a function of the theoretical range resolution $c/B$ and the corresponding operating bandwidth $B$, i.e., the bandwidth allocated for sensing. The resolution represents the ability to differentiate between two closely spaced targets in range. \textcolor{black}{This analysis outlines the bandwidth requirements for future real-time implementations; therefore, it is} relevant for real-time human body localization in ISAC scenarios, where only a portion of the available band can be reserved for communication or sensing. 
Error bars in Fig.~\ref{range_resol} indicate the standard deviation of the positioning error across different body locations. A significant increase in positioning uncertainty is observed for frequency bands below $7$~GHz, corresponding to a range resolution of approximately $4$~cm. This dimension is comparable to the size of human body parts, while positioning accuracy is further affected by ambiguity errors due to the use of a single radio link and the limited number of MPCs. A joint analysis of the ADPP from multiple links can help reduce such uncertainty.
\begin{figure*}
	\centering \includegraphics[scale=0.42]{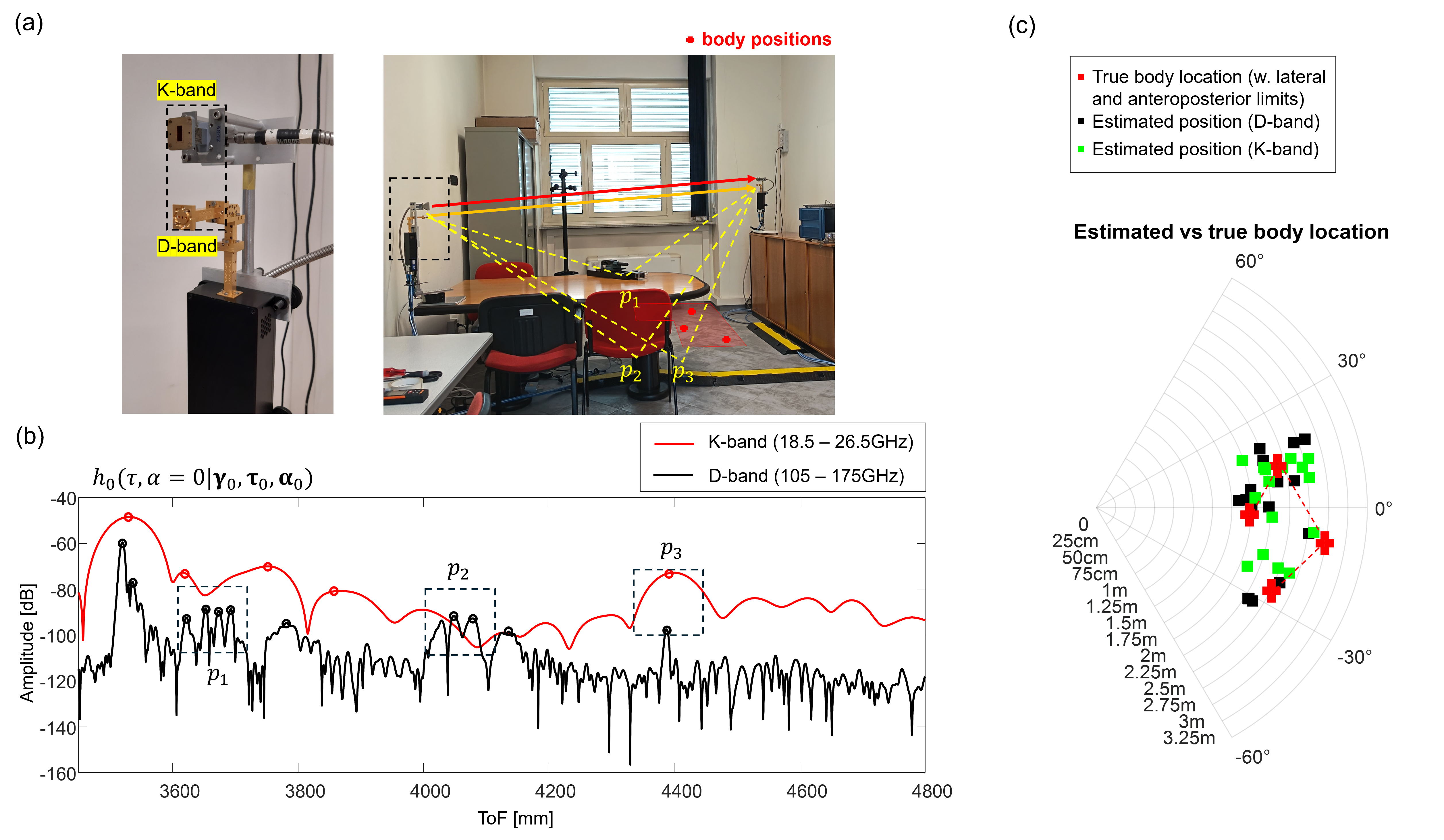} \protect\caption{\label{fig:benchmark} \textcolor{black}{D-band and K-band benchmark analysis: (a) experimental setup with co-located K-band ($18.5$--$26.5$~GHz) and D-band antennas under the same acquisition mechanism; (b) comparative CIR (reflectivity) [dB] $h_{0}(\tau,\alpha=0 \mid \boldsymbol{\gamma}_{0}, \boldsymbol{\tau}_{0}, \boldsymbol{\alpha}_{0})$ in the environment without the blocker; (c) positioning accuracy comparison for a subject located at $4$ predefined ROIs, showing ground truth (red) and estimated positions (black and green) for both bands.}}
\end{figure*}

\begin{table*}[tp]
\color{black}
\caption{\label{tab:parametersK} Vector network analyzer settings and antenna specifications at K-band}
\centering
\color{black}
\begin{tabular}{ll ll ll}
\hline\hline
\multicolumn{2}{c}{\textbf{Vector network analyzer settings}}
& \multicolumn{2}{c}{\textbf{TX antenna}} 
& \multicolumn{2}{c}{\textbf{RX antenna}} \\
\hline
\textbf{Start--stop frequency} & 18.5--26.5~GHz & \textbf{Type}        & Open-ended waveguide          & \textbf{Type}        & Pyramidal horn \\
\textbf{Frequency step}        & 1~MHz     & \textbf{FHPBW}       & $115^\circ$   & \textbf{FHPBW}       & $20^\circ$   \\
\textbf{Resolution bandwidth}  & 1~kHz        & \textbf{Polarization} & Horizontal   & \textbf{Polarization} & Horizontal     \\
\textbf{TX power}              & 0~dBm       & \textbf{Gain}        & 6.5~dBi       & \textbf{Gain}        & 20~dBi       \\
\hline\hline
\end{tabular}
\color{black}
\end{table*}

\begin{table}[tp]
    \caption{\textcolor{black}{\label{pos_acc_DK}Positioning accuracy in terms of RMSE for D-band and K-band 4 selected body locations}}
    \centering
    \begin{tabular}{lcc}
    \hline\hline
    \textbf{True body location (cm)} & \textbf{RMSE D-band} & \textbf{RMSE K-band} \\
    \textbf{} & \textbf{(cm)} & \textbf{(cm)} \\
    \hline
    $(x,y)=(204.4, 18.9)$  & $17.7$ & $84.2$ \\
    $(x,y)=(236.1,-47.7)$  & $30.7$ & $40.2$ \\
    $(x,y)=(304.1,49.9)$  & $32.7$ & $97.1$ \\
    $(x,y)=(239.4,117.2)$ & $35.2$ & $63.8$ \\
    \hline\hline
    \end{tabular}
\end{table}

\subsection{Benchmark analysis: K-band vs D-band}

To quantitatively assess the practical advantage of the proposed framework, we conducted a benchmark study with an equivalent K-band system operating at $18.5$--$26.5$~GHz. The objective is to evaluate differences in spatial resolution, MPC separability and localization accuracy under a different and more challenging indoor environment, as depicted in Fig.~\ref{fig:benchmark}. The comparative analysis of the two bands adopts a similar acquisition setup,  environment, and processing conditions.

The benchmark configuration is illustrated in Fig.~\ref{fig:benchmark}(a). The K-band and D-band antennas are co-located in the same indoor environment to ensure consistent geometry and multipath structure. The systems simultaneously operate under the same automated spatial probing mechanism described in Section~\ref{sec:setups}, including identical angular scanning and data acquisition procedures. TX and RX antenna specification are chosen to be similar to the D-band configuration with marginal deviations in therms of gain and FHPBW (see Table~\ref{tab:parameters}). A human subject is positioned at $4$ predefined ROIs within the sensing region, namely crossing the LOS path and surrounding the desk. This setup enables a direct comparison of localization performance across bands. The same MPC extraction, association, and classification pipeline is applied to both datasets (see Section~\ref{sec:Channel-Impulse-Response}), while the only adaptation concern the effective operational bandwidth. 

Fig.~\ref{fig:benchmark}(b) illustrates the qualitative differences in the MPC structure. The figure compares the CIR responses \eqref{eq:empty} for the environment without the blocker at $\alpha = 0$ and the corresponding MPCs (circle markers). The D-band CIR (black) exhibits sharper and more separable peaks, reflecting the improved temporal resolution and reduced ambiguity in path identification. The LOS and secondary MPC peaks are $50$ dB and $20$ dB above the background level, respectively. On the other hand, the K-band CIR (red) shows broader and less distinguishable MPC contributions. This is due to both the narrower frequency band (8 GHz, which widens the corresponding contributions in the impulse response) and the lower path loss that provides a higher overall scattering background. The D-band representation provides a more detailed and geometrically interpretable description of the scene. For example, in the considered indoor scenario, the D-band system is able to effectively resolve reflections originating from dominant scatterers such as: the metallic base and the desk surface ($p_1$: $\tau\simeq365-370$ cm), the metallic support columns at the base of the table ($p_2$: $\tau\simeq405-410$ cm) and the floor ($p_3$: $\tau=439$ cm). As shown in the following this property is essential for reliable passive localization.

Under the considered D-band experimental conditions, the free-space path loss is approximately $86$~dB. The expected measured  $S_{21}$ in LOS conditions is again approximately $-60$~dB when accounting for the antenna gains reported in Table~\ref{tab:parameters}. Given a noise floor of about $-85$~dB on the frequency-domain $S_{21}$ trace for the adopted VNA configuration (transmitted power $13$~dBm and resolution bandwidth $1$~kHz), the LOS component exhibits an Signal-to-Noise Ratio (SNR) of roughly $25$~dB for each frequency sample. Such a value is sufficient to achieve the clear detection of the various scattering contributions that is visible in the CIR response reported in Fig.~\ref{fig:benchmark}(b). 

Fig.~\ref{fig:benchmark}(c) reports the estimated body positions for $4$ selected locations surrounding the desk: the K-band (green markers) and D-band (black markers) results are compared against the ground-truth body contour (red markers). Table~\ref{pos_acc_DK} summarizes the positioning RMSE for the corresponding locations and their true coordinates. The D-band configuration consistently provides improved positioning accuracy, yielding in most cases more than a twofold improvement in RMSE compared to the K-band system. In contrast, the K-band estimates exhibit larger spatial spread and increased bias, particularly for target locations with offsets $|y| > 50$cm from the LOS.
The higher bandwidth of the D-band system ($70$~GHz) enables millimeter-scale delay resolution, allowing closely spaced MPCs to be resolved and individually classified. This directly enhances the reliability of the MPC birth--death analysis and improves the geometric interpretation of body-induced scattering. Conversely, the narrower bandwidth of the K-band system ($8$~GHz) limits temporal resolution, leading to partial overlap of MPCs and reduced capability to isolate newly formed reflections associated with the presence of the subject. By comparing the K-band results with those reported in Fig.~\ref{range_resol} and considering an equivalent effective bandwidth, i.e., $B=7$~GHz, it can be observed that a performance gap still persists. This residual improvement is attributed to the shorter wavelength of the D-band system, which enhances angular discrimination and sensitivity to fine geometric features of the scene.

Overall, the benchmark analysis confirms that while the same unified processing pipeline can be applied at both frequency bands, the sub-THz D-band significantly enhances MPC discrimination and positioning performance. This comparison further supports the necessity of high-resolution MPC modeling for passive localization and justifies the adoption of sub-THz frequencies for single-link RF sensing applications.

\section{Conclusions\label{sec:Conclusions}}

The paper reports an analysis of human-scale blockage and scattering in the unexplored sub-THz band (D-band, 105--175~GHz), with emphasis on RF sensing and localization. The work develops and validates a comprehensive and unified framework to accurately model the complex interactions between sub-THz radio waves and a variety of blockers/scatterers, including objects of different materials and the human body. The proposed framework moves beyond conventional approaches by integrating both blockage (attenuation due to shadowing) and scattering effects as interconnected phenomena particularly at sub-THz frequencies.

A key contribution of this work is the design of a signal processing framework that classifies angle--delay MPCs into distinct categories: newly formed, attenuated/enhanced, or suppressed by blockage events. Based on extensive indoor measurements, the study demonstrates how targets and blockers influence the electromagnetic field and govern the birth--death process of MPCs. Test objects of different sizes and materials (metal and paper) as well as human subjects were considered. Results show that blocker position and dimensions can be effectively reconstructed from the time-varying MPCs of a single radio link. Furthermore, discrimination between materials with markedly different electromagnetic properties is achieved. Results show that body micro-movements of 5~cm are detectable, offering improved accuracy compared to conventional RF sensing at cm-scale wavelengths (2.4--6~GHz). \textcolor{black}{The benchmark analysis between K-band ($18.5$--$26.5$~GHz) and D-band measurements further highlights the advantages of sub-THz sensing, showing that the higher bandwidth and shorter wavelength of the D-band enable improved multipath discrimination and localization accuracy.} These findings highlight the potential of sub-THz bands for high-resolution occupancy detection and passive RF sensing applications.

Building on the presented analysis, the time-varying characteristics of MPC birth--death processes are expected to be crucial for classifying various body activities and scenarios.
While human body effects were considered, future work will address dynamic interactions with moving subjects, including the tracking  of postures, gestures, gaits, and motion patterns. \textcolor{black}{The proposed results demonstrate the effectiveness of the framework using a single sub-THz link. On the other hand, we clearly expect significant performance improvements when extending the approach to multiple links as in classical RF holography/tomography approaches.}

\vspace{-1cm}
\begin{IEEEbiographynophoto}{Stefano Savazzi }(Senior Member, IEEE) is a Senior Researcher at Consiglio Nazionale delle Ricerche (CNR), the Institute of Electronics, Computer and Telecommunication Engineering (IEIIT). He received the M.Sc. degree and the Ph.D. degree (Hons.) in ICT from the Politecnico di Milano, Italy, in 2004 and 2008, respectively, and joined CNR in 2012. He was a Visiting Researcher with Uppsala University, in 2005 and University of California at San Diego in 2007. He has coauthored over 140 scientific publications (Scopus). His current research interests include distributed signal processing, distributed machine learning and networking aspects for the Internet of Things, radio localization and vision technologies. Dr. Savazzi was the recipient of the 2008 Dimitris N. Chorafas Foundation Award. He is principal investigator for CNR in the Horizon EU projects Holden, TRUSTroke and the Doctoral Network SMARTTEST. He is also serving as Associate Editor for Frontiers in Communications and Networks, Wireless Communications and Mobile Computing, Personal Wireless Communications and Sensors. 
\end{IEEEbiographynophoto}
\vspace{-1.0cm}
\begin{IEEEbiographynophoto}{Fabio Paonessa }is a Researcher at the Institute of Electronics, Information and Telecommunications Engineering (IEIIT) of the Italian National Research Council (CNR). He received the B.S. and M.S. in biomedical engineering and the Ph.D. in electronics engineering from Politecnico di Torino, Turin, Italy, in 2008, 2010, and 2017, respectively. From 2011 to 2012, he was a Research Assistant at Politecnico di Torino, working on electronic systems for therapeutic ultrasound applications. In 2013, he joined the Applied Electromagnetics Group of the CNR-IEIIT. His research focuses on UAV-based characterization of antenna systems, including low-frequency radio astronomy arrays, Q-band polarimeters, X-band radar systems, and wireless sensor network nodes. His interests also include the design of microwave waveguide passive components and joint communication and sensing at sub-terahertz frequencies. He has authored or coauthored more than 25 journal articles and 50 conference papers and has contributed to several national and international research projects.
\end{IEEEbiographynophoto}
\vspace{-1cm}
\begin{IEEEbiographynophoto}{Sanaz Kianoush} (sanaz.kianoush@cnr.it) is a Senior Researcher at the Consiglio Nazionale delle Ricerche (CNR), Institute of Electronics, Computer and Telecommunication Engineering (IEIIT) in Italy. She received her Ph.D. degree in Electronic Engineering from the University of Pavia in 2014 and joined CNR-IEIIT as a postdoctoral researcher.In 2022, she was appointed as Researcher at CNR-IEIIT and was promoted to Senior Researcher in 2023.In 2018, she was a visiting researcher at Aalto University. In 2008-10, she was the lecturer at Azad University Sary (Iran). She won CNR-IEIIT Young Research Award (PreGio) in 2018. She has authored or coauthored more than 60 journal articles and conference papers. Her research interests include statistical signal processing, sensor fusion in communication systems, machine learning and federated learning, as well as the development of smart radio environments for healthcare, human monitoring, and human machine collaborative applications.   
\end{IEEEbiographynophoto}
\vspace{-1cm}
\begin{IEEEbiographynophoto}{Alessandro Nordio} is a Senior Researcher at the Istitute of Electronics, Information Engineering and Telecommunications of the Italian National Research Council (CNR-IEIIT). In 2002 he received the Ph.D. in Telecommunications from EPFL, Lausanne, Switzerland. From 2002 to 2009 he was with the Electronic Department of Politecnico di Torino, Italy. He published more than 50 journal articles and tens of conference papers. His research interests are in the field of signal processing, smart radio environments, and 6G and beyond communications.
\end{IEEEbiographynophoto}
\vspace{-1cm}
\begin{IEEEbiographynophoto}{Giuseppe Virone } was born in Turin, Italy, in 1977. He received the Laurea degree (summa cum laude) in electronic engineering, and the Ph.D. degree in electronics and communication engineering from the Politecnico di Torino, Turin, Italy, in November 2001 and 2006, respectively. He is currently a Director of  Researcher at the Istituto di Elettronica e di Ingegneria Informatica e delle Telecomunicazioni (IEIIT), Italian National Research Council (CNR), Turin. He joined IEIIT as a Research Assistant in 2002. He coordinated more than 25 scientific projects/research units/work-packages funded by both the industry and other scientific research organizations and joined more than 30 research projects as a collaborator. He authored 60 journal papers, 120 conference papers, and three European patents. His activities concern the design, numerical analysis, and characterization of microwave, millimeter passive components and sub-THz components for feed systems, antenna arrays, frequency selective surfaces, compensated dielectric radomes for communication and sensing applications. He is also active in advanced signal processing and propagation measurements for smart radio environments. 
\end{IEEEbiographynophoto}

\end{document}